\documentclass[journal]{IEEEtran}
\usepackage{graphicx}
\usepackage{amssymb,amsmath}
\usepackage{amsfonts}
\usepackage{epstopdf}
\usepackage{mathrsfs}
\usepackage{array}
\usepackage{subfigure}
\DeclareMathAlphabet{\mathpzc}{OT1}{pzc}{m}{it}
\usepackage{cite}
\usepackage[ruled,vlined]{algorithm2e}
\usepackage[para,online,flushleft]{threeparttable}
\usepackage{algpseudocode}

\begin{document}
\title{Neural Networks based EEG-Speech Models}
\author{Pengfei~Sun,
        and~Jun~Qin,}     
\maketitle
\begin{abstract}
In this paper, we propose an end-to-end neural network (NN) based EEG-speech (NES) modeling framework, in which three network structures are developed to map imagined EEG signals to phonemes. The proposed NES models incorporate a language model based EEG feature extraction layer, an acoustic feature mapping layer, and a restricted Boltzmann machine (RBM) based the feature learning layer. The NES models can jointly realize the representation of multichannel EEG signals and the projection of acoustic speech signals. Among three proposed NES models, two augmented networks utilize spoken EEG signals as either bias or gate information to strengthen the feature learning and translation of imagined EEG signals. Experimental results show that all three proposed NES models outperform the baseline support vector machine (SVM) method on EEG-speech classification. With respect to binary classification, our approach achieves comparable results relative to deep believe network approach. 
\end{abstract}
\IEEEpeerreviewmaketitle

\section{Introduction}
\IEEEPARstart{H}{uman}-computer-interfaces (HCIs) involve electroencephaloggraphy (EEG) based communication \cite{toda2008statistical}. By translating EEG patterns (i.e., user intent) into control signals, HCIs are used to operate machines such as computers or assistive devices. As one of major brain activities, imagined speech can be a promising carrier to implement intuitive HCIs. Recently, automatic speech recognition technique has facilitated our lives by providing a new approach of communication with machines. Previous research \cite{di2015low} demonstrates that it is feasible to use EEG to recognize human speech, and to broaden the boundary of human-machine perception.  

The key issue of EEG based speech recognition is to model the relationship between EEG signals and speech parameters. Though various studies \cite{lotte2007review, gao2014visual} demonstrate that speech activities can be represented by EEG data, it is still very difficult to correlate the two types of signals. The major challenges  include: 1) EEG signals are easily contaminated by artifacts produced by muscle movements, and 2) the relevant brain activities (e.g., prior or posterior emotions) significantly affect the presenting forms of speech related EEG signals \cite{herff2016automatic}. Therefore, the observed EEG signals are generally considered as the mixture of multiple sources, and speech associated EEG signals are masked by other brain activities. The features (e.g., mean, kurtosis and entropy) used in conventional EEG-Speech classification techniques, may not yield deep representation of speech related EEG data. 

Another practical issue for EEG-speech models is the multimodal data (i.e., EEG signals and acoustic signals) fusion. One popular approach is to symbolize each acoustic phoneme or word, and then pair the EEG signals with corresponding label \cite{matsumoto2014classification}. This procedure reduces high dimensional acoustic signals into low dimensional language label, and as a result, cannot reveal the intrinsic physical responses of brain (i.e., EEG signals) evoked by acoustic stimulus. Therefore, it may lose accuracy and resolution of the feature space provided by EEG scalp distribution. Moreover, in the higher dimensional acoustic feature space, it could be much easier to construct unique mapping relationships between EEG-Speech signal pairs.  

To address these issues, we propose an end-to-end neural networks (NN) based EEG-speech (NES) modeling framework. The basic structure of this NES framework consists of a linear EEG feature extraction layer, a restricted Boltzmann machine (RBM) based feature learning layer, and a linear speech feature projection layer. By introducing a language model based feature extraction layer, the NES framework can utilize the correlations among different channels of EEG signals, and obtain deep feature representation. The succeeded RBM layer can strengthen EEG feature learning and suppress the artifacts of EEG signals accordingly. The speech projection layer devotes to achieve the multimodal fusion. The obtained EEG features will be translated into speech feature space. In detail, three NES models are developed, including an imagined EEG-speech (NES-I) model, a biased imagined-spoken EEG-speech (NES-B) model, and a gated imagined-spoken EEG-speech (NES-G) model. NES-I model maps imagined EEG signals to speech signals. NES-B and NES-G models incorporate spoken EEG signals as bias and gate information, respectively, to strengthen the feature learning and translation of imagined EEG signals. Especially, NES-G model applies a factored RBM training approach \cite{sun2016enhanced} to implement the learning process.
\begin{figure*}[hbt]
\vspace{-4mm}
\centerline{\includegraphics[scale=1]{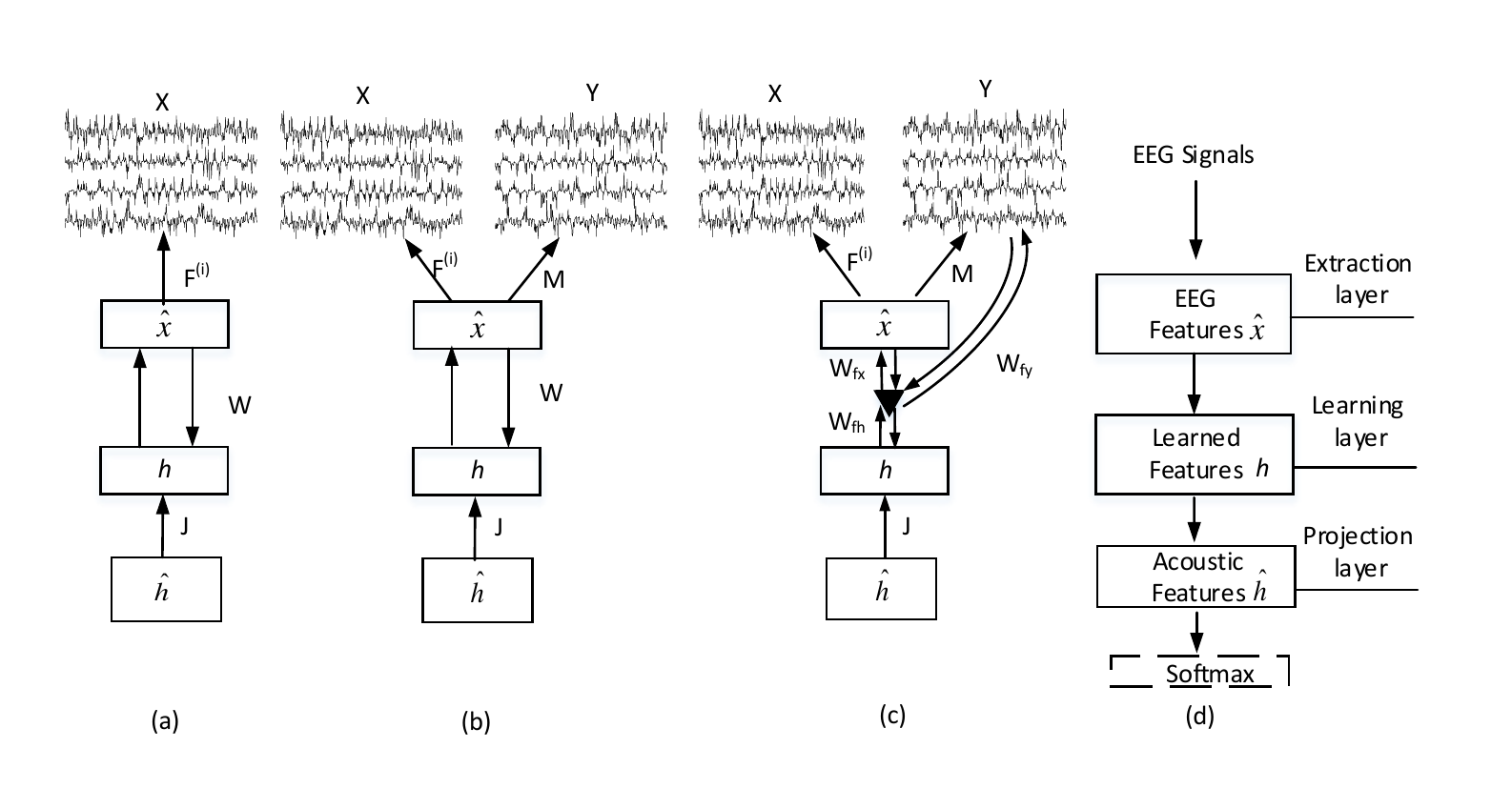}}
\vspace{-3mm}
\caption{The schematic diagram of three proposed neural network based EEG-speech models, (a)NES-I, (b)NES-B, and (c)NES-G, respectively. The arrow in the figure represents the training flow. The general NES framework is described in (d), and the added softmax layer is specifically applied for classification.}
\label{threemodels}
\end{figure*}

\subsection{Related Work}
In EEG based speech classification and recognition fields, a variety of studies have been conducted. In the branch of imagined speech recognition, various voice related potentials, i.e., vowels, phonemes, syllables, and whole-word \cite{suppes1997brain, dasalla2009single, d2009toward, lopez2012auditory}, are used as speech labels. Porbadnigk $et$ $al.$ \cite{porbadnigk2009eeg} applied an HMM to classify EEG signals associated with consecutive words, and the limited performance revealed that the order of words significantly affected the results. In other studies, imagined EEG signals have also been used to decode variety of languages \cite{dasalla2009single, hausfeld2012pattern, wang2013analysis}. Recently, Zhao and Rudzicz \cite{zhao2015classifying} investigated spoken EEG signals based speech classification by utilizing a deep belief network. In their model, the inputs were still shallow features gathered by conventional techniques. Di Liberto $et$ $al.$ \cite{di2015low} and O'Sullivan $et$ $al.$ \cite{o2015attentional} both worked on spoken EEG translation and devoted on directly mapping to the acoustic features of speech (i.e., envelope or spectrogram). Yoshimura $et$ $al.$ incorporated fMRI data as a hierarchical prior for EEG feature extraction by using a variational Bayesian method \cite{yoshimura2016decoding}. Their results showed that adding extra imagined speech information can boost the EEG based recognition. In this study, the similar idea has been incorporated into our NES models by combining both spoken and imagined EEG signals.
  
In the fields of multimodal representation learning, several deep learning methods have been introduced to successfully extract the intrinsic structures from multiple modalities. Ngiam $et$ $al.$ \cite{ngiam2011multimodal} learned features from audio and video by utilizing deep autoencoders. Srivastava $et$ $al.$ \cite{srivastava2012multimodal} developed a model of images and text relying on the multimodal deep Boltzmann machine. More recently, Socher $et$ $al.$\cite{socher2014grounded} and Frome $et$ $al.$ \cite{frome2013devise} proposed methods for mapping images into a text representation space. The idea of domain transformation in their work is quite similar to our approach that we project both EEG and speech signals into the corresponding feature spaces. Specifically, the feature space of EEG signals is learned from a context model derived from the concept of log-bilinear language model \cite{mnih2007three}. In addition, Kiros $et$ $al.$ \cite{kiros2014multimodal} proposed a 3-way network structure to enforce the joint image-text learning. By adding extra gate branch, such network structure can efficiently incorporate the underlying correlations. In our proposed models, the gated NN are utilized to incorporate the relationships between imagined and spoken EEG signals.  

\section{EEG-Speech models}
\subsection{Network Structures of Three Models}
Based on the basic structure illustrated in Fig.~\ref{threemodels}(d), three NES models are developed to map imagined EEG signals to speech signals. In NES-I model (as shown in Fig.\ref{threemodels}(a)), feature extraction layer (top) and projection layer (bottom) are both connected to the middle layer, which is a RBM based feature learning layer. The input and output of NES-I model are imagined-EEG signals and phonemes, respectively. NES-B model (as shown in Fig.\ref{threemodels}(b)) augments NES-I model by incorporating  spoken EEG signals to bias imagined EEG signals at the input layer. As shown in Fig.\ref{threemodels}(c), NES-G model introduces the multiplicative interaction between spoken EEG and imagined EEG signals. In NES-G model, spoken EEG signals are not only used as bias, but also as a conditional branch to gate imagined EEG signals. In both NES-B and NES-G models, spoken EEG signals are applied to demonstrate the possibility of using correlated information to strengthen imagined EEG signal recognition. 

\subsubsection{EEG Feature Extraction and Mapping}
\begin{figure}[hbt]
\vspace{-7mm}
\centerline{\includegraphics[scale=1]{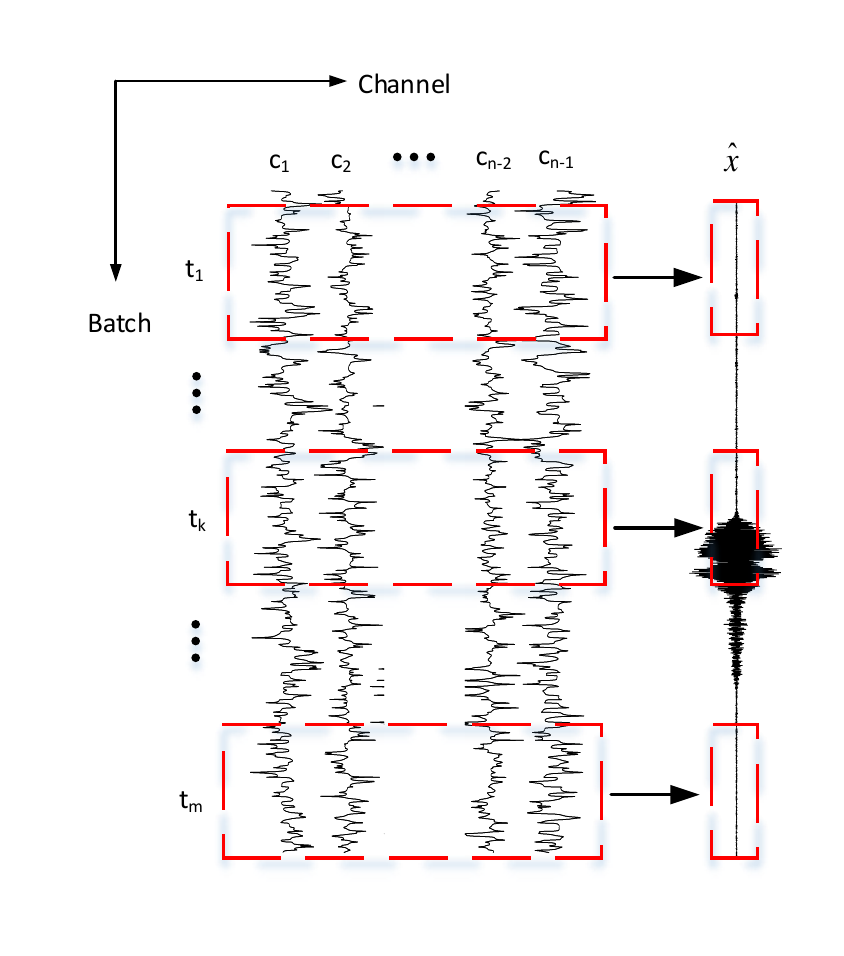}}
\vspace{-8mm}
\caption{The schematic diagram of joint-context EEG-Speech tuple. The horizontal axle is the channel label and the vertical axle is time batch.}
\vspace{-3mm}
\label{EEG_Speech}
\end{figure}

A conventional imagined EEG-speech model is implemented based on the pre-extracted EEG features. Comparatively, by adding a EEG linear transform layer as shown in Fig.~\ref{threemodels}, NES models integrate the feature extraction into networks of feature learning and mapping. In EEG feature extraction layer, the concept of language model originally developed by Kiros $et$ $al.,$ \cite{kiros2014multimodal}, Mnih and Hinton \cite{mnih2007three} is borrowed. This language model assumes that the semantic sequences can be used to predict the next word because of the existed correlation. Similarly, we regard that a unique imagined EEG pattern corresponding to each individual speech segment exists, and is also bundled with a series of EEG signals. The principle is illustrated in Fig.~\ref{EEG_Speech}. Each EEG channel $c_{i}$ is treated as the context of the rest channels $c_{j, j \neq i}$. Therefore, the EEG signals at the batch $t_{k}$ are considered as a tuple. Accordingly, the n-1 channel EEG signals ${x_{c_{1},t_{k}},\cdots,x_{c_{n-1},t_{k}}}$ are followed by corresponding speech feature $\hat{x}_{t_{k}}$. 

Mathematically, EEG signals in each channel $c_{i}$ is represented as a $D$-dimensional real-valued vector $\mathbf{x}_{c_{i}}$, and the linear transformation can be written as  
\begin{equation}
\hat{\mathbf{x}} = \sum_{i=1}^{n-1}\mathbf{x}_{c_{i}}\mathbf{F}^{(i)}
\end{equation}   
where $\mathbf{F}^{(i)}$, $i=1,\cdots,n-1$ are $D\times D$ feature parameters matrices. $\mathbf{F}^{(i)}$ works to transform the $\mathbf{x}_{c_{i}}$ into the feature space $\mathcal{W}^{D}$, and $\hat{\mathbf{x}}$ as the summation of the feature vectors from all EEG channels represents the predicted 'word'. 

In order to obtain the higher level feature representation $h$ shown in Fig.~\ref{threemodels}, $\hat{\mathbf{x}}$ then is fed into the RBM for unsupervised feature learning. To realize the modal fusion between learned EEG features $h$ and speech features $\hat{\mathbf{h}}$, a linear projection is implemented to transform $h$ into the feature space of $\hat{\mathbf{h}}$. This linear transformation can be described as $\hat{\mathbf{h}} = \mathbf{h} \mathbf{J}$. The transformation matrix $\mathbf{J}$ is $K\times L$, where $K$ is the dimension of speech feature space and $L$ is the dimension of phoneme vector. Both $F^{(i)}$ and $J$ can be updated during the NES models training.

\subsubsection{Imagined EEG-Speech Model (NES-I)}
For the basic NES model, NES-I (as shown in Fig.\ref{threemodels}(a)) utilizes a typical RBM as the middle layer to extract EEG features. To define the middle layer RBM model for the distribution of the acoustic feature $\hat{x}$, given the context $x_{c_{1:n-1}}$, an energy function must be specified for a joint configuration of the visible and hidden units. Briefly, we start by specifying the energy function:
\begin{equation}
E(\mathbf{\hat{x}},\mathbf{h};c_{1:n-1}) = -\hat{\mathbf{x}}^{T} W\mathbf{h}-\mathbf{b}_{h}^{T}\mathbf{h}-\mathbf{b}_{\hat{x}}^{T}\hat{\mathbf{x}}
\end{equation}
The vector $\mathbf{b}_{h}$ contains the biases for hidden units, while vectors $\mathbf{b}_{\hat{x}}$ contain biases for EEG signals. In the context of a feed-forward NN, the weight between the EEG features $\hat{\mathbf{x}}$ and $\mathbf{h}$ is the representation transformation matrix $\mathbf{W}$. Considering that logistic units may not be sufficient to represent EEG feature vector $\mathbf{\hat{x}}$, the energy function can be modified as a Gaussian RBM:
\begin{equation}
E(\mathbf{\hat{x}},\mathbf{h};c_{1:n-1}) = \frac{(\hat{\mathbf{x}}-\mathbf{b}_{\hat{x}})^{T}(\hat{\mathbf{x}}-\mathbf{b}_{\hat{x}})}{2 \mathbb{\sigma}^{2}} -\hat{\mathbf{x}}^{T} W\mathbf{h}-\mathbf{b}_{h}^{T}\mathbf{h}
\label{energyf}
\end{equation}

The conditional distribution of the hidden configuration $\mathbf{h}$ is defined as 
\begin{equation}
P(\mathbf{h}|\mathbf{\hat{x}} ,c_{1:n-1}) = \frac{1}{Z_{c}}exp(-E(\mathbf{\hat{x}}, \mathbf{h};x_{c_{1:n-1}}))
\end{equation} 
\begin{equation}
Z_{c_{1:n-1}} = \sum_{\mathbf{\hat{x}}, \mathbf{h}}exp(-E(\mathbf{\hat{x}},\mathbf{h};x_{c_{1:n-1}})),
\end{equation}
where $Z_{c_{1:n-1}}$ is a context-dependent normalization term summed over $\mathbf{\hat{x}}$ and $\mathbf{h}$. Since there is no connection between neurons in the same layer, inferences of the $k$th hidden and $j$th visible units can be performed as 
\begin{equation}
p(h_{k}=1|\mathbf{\hat{x}}) = S(\Delta E_{k})
\label{eqhp}
\end{equation}
\begin{equation}
p(\hat{x}_{j} = x|\mathbf{h}) = N(x|\Delta E_{j},\sigma_{j}^{2})
\label{eqvp}
\end{equation}
where $N(\cdot|\mu, \sigma^{2})$ denotes the Gaussian probability density function with mean $\mu$ and standard deviation $\sigma$. $S(\cdot)$ is the sigmoid activation function. $\Delta E_{k}$ and $\Delta E_{j}$ are the overall inputs of the $k$th hidden unit and $j$th visible unit, respectively \cite{yamashita2014bernoulli}. Unlike existing EEG based speech classification techniques, the proposed NES-I model learns the deep feature representations relying on its top-bottom NN structure. The output $\mathbf{h}$ can be projected as phoneme $\mathbf{\hat{h}}$. Learning process is a combination of standard back propagation based supervised training and contrast divergence based unsupervised training. 

\subsubsection{Biased Imagined-spoken EEG-Speech Model (NES-B)} 
Supposing that along with each training tuple of imagined EEG signals $(x_{c_{1}},\cdots,x_{c_{n-1}})$, there exists an associated vector $\mathbf{y}$ $\in \mathcal{R}^{M}$ corresponding to the feature representation of the modality to be conditioned on, such as spoken-speech related EEG signals. By utilizing the spoken EEG signals to strengthen feature extraction of imagined EEG signals, NES-B model is a straightforward extension of the NES-I model as described in Fig.~\ref{threemodels}(b). The additive bias $\mathbf{y}$ is added to suppress the artifacts, and accordingly the hidden feature $\mathbf{\hat{x}}$ can be described as 
\begin{equation}
\hat{\mathbf{x}} = \left(\sum_{i=1}^{n-1}\mathbf{x}_{c_{i}}\mathbf{F}^{(i)}\right)+\mathbf{y}\mathbf{M}^{m}
\end{equation} 
where $\mathbf{M}^{m}$ is a $ M \times D$ feature matrix that projects the spoken EEG signals into the EEG feature space $\mathcal{W}^{D}$. This model has similar training procedure as NES-I model: given acoustic feature representation $\hat{\mathbf{x}}$, we can obtain the conditional distribution $p(\mathbf{h}|c_{1:n-1}, \hat{\mathbf{x}})$ following (\ref{eqhp}). Unlike the constant bias, the spoken EEG branch can be viewed as a locally changed variable. After linear transformation, the bias term $\mathbf{y}\mathbf{M}^{m}$ is supposed to amplify the speech related components, while reduces the deviation caused by artifacts. The linear transformation $\mathbf{M}$ can be obtained when back propagation training is implemented.

\subsubsection{Gated Imagined-spoken EEG-Speech Model (NES-G)} 
As an advance extension of NES-B model, NES-G retains the bias projection developed in NES-B, while applies the projected spoken EEG term $\mathbf{y}\mathbf{M}^{m}$ to multiply with hidden speech features $\mathbf{\hat{x}}$. In other words, the proposed NES-G model uses spoken EEG signals to modulate the input layer, and this modulation procedure can be regarded as a gated relationship. The gated RBMs in the middle layer allow hidden units to model the transition between successive EEG batches.

By doing this, $\mathbf{W}$ becomes a tensor. Therefore, $\mathbf{W}$ requires $D \times M \times K$ parameters, where $K$ is the dimension of hidden layer. Correspondingly, the energy function (\ref{energyf}) can be rewritten as 
\begin{equation}
\begin{split}
&E(\mathbf{y},\mathbf{h},\mathbf{\hat{x}}) = -\sum_{k}b_{k}h_{k}+\sum_{j}\frac{(y_{j}-b_{j})^{2}}{2\sigma_{j}^{2}}+\\ 
&\sum_{i}\frac{(\hat{x}_{i}-b_{i})^{2}}{2\sigma_{i}^{2}}
- \sum_{ijk}w_{ijk}\frac{x_{i}}{\sigma_{i}} \frac{y_{j}}{\sigma_{j}} h_{k}
\end{split}
\label{eqfactor}
\end{equation}
The 3D tensor $w_{ijk}$ is impractical for general network training due to high volume of parameters. A solution is to factor $\mathbf{W}$ into three lower-rank matrices $\mathbf{W}_{fx}$, $\mathbf{W}_{fy}$ and 
$\mathbf{W}_{fh}$. For each acoustic feature, $\mathbf{h}$ can specify its own hidden and gated weight matrix. In details, $(\mathbf{W}_{fx})^{T}$, and $\mathbf{W}_{fy}$ denote the imagined and spoken EEG embeddings. Given a EEG feature representation $\hat{\mathbf{x}}$, the factor outputs are 
\begin{equation}
f = (\mathbf{W}_{fx}\hat{\mathbf{x}})\cdot (\mathbf{W}_{fy}\mathbf{y})
\label{multioutput}
\end{equation}
where $\cdot$ is an element-wise product. Equation (\ref{multioutput}) helps to understand the advantage of NES-G model: the multiplication is beneficial to strengthen the highly correlated components between imagined EEG and spoken EEG signals. As a result, it can greatly suppress those EEG features with less similarity. Comparing with NES-B model, NES-G model is more efficient to catch underlying features of imagined and spoken EEG signals. 

\subsection{Gated restricted Boltzmann machine learning} 
To efficiently train NES-G model, the three way tensor $\mathbf{W}$ in (\ref{eqfactor}) should be factored into decoupled matrices. In this section, for generalization, $\mathbf{\hat{x}}$ is noted as $\mathbf{x}$. Accordingly, the energy function can be given as \cite{sun2016enhanced}:
\begin{equation}
\begin{split}
&E(\mathbf{y},\mathbf{h},\mathbf{x}) = -\sum_{k}b_{k}^{h}h_{k}+\sum_{j}\frac{(y_{j}-b_{j}^{y})^{2}}{2\sigma_{j}^{2}}+ \sum_{i}\frac{(x_{i}-b_{i}^{x})^{2}}{2\sigma_{i}^{2}}\\
&- \sum_{f}\left(w_{if}^{x}\sum_{i}\frac{x_{i}}{\sigma_{i}}\right) \left(\sum_{j} w_{jf}^{y}\frac{y_{j}}{\sigma_{j}} \right) \left( \sum_{k}w_{kf}^{h}h_{k}\right)
\end{split}
\label{eqfactor1}
\end{equation}
By noting 
\begin{equation}
f_{f}^{x}= \sum_{i=1}^{D}w_{if}^{x}\frac{x_{i}}{\sigma_{i}}, \quad  f_{f}^{y}= \sum_{j=1}^{M}w_{jf}^{y}\frac{y_{j}}{\sigma_{j}}, \quad f_{f}^{h}= \sum_{k=1}^{K}w_{kf}^{h}h_{k}
\end{equation}
where $i$, $j$ and $k$ index input, visible and hidden units, respectively. $x_{i}$ and $y_{j}$ are Gaussian units, and $h_{k}$ is the binary state of the hidden unit $k$. $\sigma_{i}$ and $\sigma_{j}$ are standard deviations associated with $x_{i}$ and $y_{j}$, respectively. The terms $b_{k}^{h}$ and $b_{j}^{y}$ represent biases of the hidden and observable units, respectively. Inferences of the $k$th hidden and $j$th visible unit can be given as
\begin{equation}
p(h_{k}=1|\mathbf{y};\mathbf{x}) = S(\Delta E_{k})
\label{eqhp1}
\end{equation}
\begin{equation}
p(y_{j} = y|\mathbf{h};\mathbf{x}) = N(y|\Delta E_{j},\sigma_{j}^{2})
\label{eqvp1}
\end{equation}
where $N(\cdot|\mu, \sigma^{2})$ denotes the Gaussian probability density function with mean $\mu$ and standard deviation $\sigma$. $S(\cdot)$ is the sigmoid activation function. $\Delta E_{k}$ and $\Delta E_{j}$ are the overall inputs of the $k$th hidden unit and $j$th visible unit, respectively. The gated RBMs allow hidden units to model the transition between successive frames, and the input units are collected directly from previous frames in many applications \cite{memisevic2010learning}.
\begin{figure}[htb]
\vspace{-4mm}
\centerline{\includegraphics[scale=0.7]{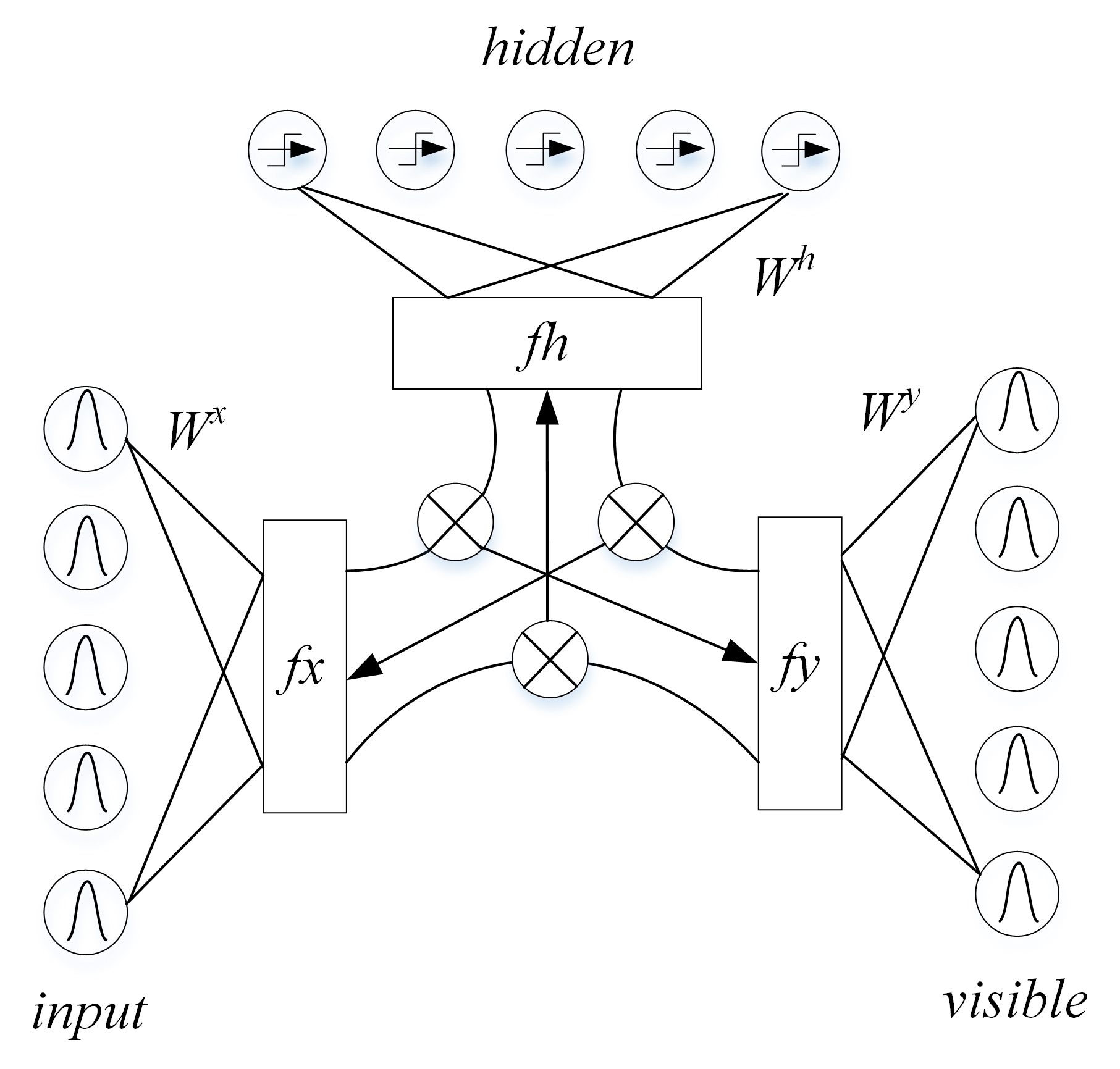}}
\vspace{-3mm}
\caption{The schematic diagram of symmetrical three-way RBMs. The three factor layers have the same size, and $\otimes$ refers to element-wise multiplication.}
\vspace{-3mm}
\label{recursiveNN}
\end{figure}

The three factor layers as shown in Fig.~\ref{recursiveNN} have the same factor size $F$, and the factor terms (i.e., $W^{x}\mathbf{x}$, $W^{y}\mathbf{x}$, and $W^{h}\mathbf{h}$) correspond to three linear filters applied to the input, visible, and the hidden units, respectively. To perform k-step Gibbs sampling in the factored model, the overall inputs of each unit in the three layers are calculated as
\begin{equation}
\Delta E_{k} = \sum_{f}w_{kf}^{h}\sum_{i}w_{if}^{x}\frac{x_{i}}{\sigma_{i}}\sum_{j}w_{jf}^{y}\frac{y_{j}}{\sigma_{j}}+ b_{k}^{h}
\label{hidden}
\end{equation} 
\begin{equation}
\Delta E_{j} = \sum_{f}w_{jf}^{y}\sum_{i}w_{if}^{x}\frac{x_{i}}{\sigma_{i}}\sum_{k}w_{kf}^{h}h_{k}+ b_{j}^{y}
\label{visible}
\end{equation}
\begin{equation}
\Delta E_{i} = \sum_{f}w_{if}^{x}\sum_{j}w_{jf}^{y}\frac{y_{j}}{\sigma_{j}}\sum_{k}w_{kf}^{h}h_{k}+b_{i}^{x}
\label{input}
\end{equation}

In (\ref{hidden})-(\ref{input}), the factor layers are multiplied element-wise (as the $\otimes$ illustrated in Fig.~\ref{recursiveNN}) through the same index $f$. These are then substituted in (\ref{eqhp})-(\ref{eqvp}) to determine the probability distributions for each of the visible and hidden units. Therefore, each speech pattern in the hidden units corresponds to a pairwise matching of input filter responses and visible filter responses. The learning procedure aims to find a set of filters that can represent the correlations of consecutive speech frames in the training data.

To train factored RBMs, one needs to maximize the average log-probability $L=$log $p(\mathbf{y}|\mathbf{x})$ of a set of training pairs $\{(\mathbf{x}, \mathbf{y})\}$. The derivative of the negative log-probability with respect to parameters $\theta$ is given as 
\begin{equation}
-\frac{\partial L}{\partial \theta} = \langle \frac{\partial E(\mathbf{y}, \mathbf{h}; \mathbf{x})}{\partial \theta} \rangle_{\mathbf{h}}-\langle \frac{\partial E(\mathbf{y}, \mathbf{h}; \mathbf{x})}{\partial \theta} \rangle_{\mathbf{h},\mathbf{y}}
\label{eqderi}
\end{equation}
where $\langle \rangle_{\mathbf{v}}$ denotes the average with respect to variable $\mathbf{v}$. In practice, Markov chain step running is used to approximate the averages in Eq.~(\ref{eqderi}). By differentiating (\ref{eqfactor}) with respect to the parameters, we get 
\begin{equation}
-\frac{\partial E}{\partial w_{kf}^{h}} = -h_{k}\sum_{i}x_{i} w_{if}^{x}\sum_{j}y_{j}w_{jf}^{y}
\label{liklyhood1}
\end{equation}
\begin{equation}
-\frac{\partial E}{\partial w_{jf}^{y}} = -y_{j}\sum_{i}x_{i}w_{if}^{x}\sum_{k}h_{k}w_{kf}^{h}
\label{liklyhood2}
\end{equation}
\begin{equation}
-\frac{\partial E}{\partial w_{if}^{x}} = -x_{i}\sum_{j}y_{j}w_{jf}^{y}\sum_{k}h_{k}w_{kf}^{h}
\label{liklyhood3}
\end{equation}
\begin{equation}
-\frac{\partial E}{\partial b_{k}^{h}} = h_{k}, \quad -\frac{\partial E}{\partial b_{i}^{x}} = x_{i}, \quad -\frac{\partial E}{\partial b_{j}^{y}} = y_{j}
\label{liklyhood4}
\end{equation}

In this study, we use the power of EEG and speech signals as the input $\mathbf{x}$ and output $\hat{\mathbf{h}}$. In order to encourage nonnegativity in three factored matrices $w_{kf}$, $w_{if}$, and $w_{jf}$, a quadratic barrier function is incorporated to modify the log probability. The objective function is the regularized likelihood as below \cite{nguyen2013learning}
\begin{equation}
- \mathcal{L}_{reg}= \mathcal{L}(\mathbf{y};\mathbf{x})- \frac{\alpha}{2}\sum_{}\sum_{}f(w)
\label{sparsity}
\end{equation}
where 
\[
    f(x)=\left\{
                \begin{array}{ll}
                  x^{2} \quad x<0\\
                  0 \qquad x\geq 0
                \end{array}
              \right.
\]

A joint unsupervised and supervised training is used to map imagined-EEG signals $\mathbf{x}$ to phonemes $\hat{\mathbf{h}}$ stepwise. The linear feature transformations as parts of the NN are also updated during the network learning. For all three NES models, the training sequence to obtain network parameters can be given in following algorithm framework
\begin{algorithm} 
\caption{Joint unsupervised-supervised training}  
 \For{iteration $\leq$ $N_{epoch}$}   
  {
    \For{Iteration $\leq$ $N_{batch}$}
    {
         \textbf{Forward} \\
         $\hat{\mathbf{x}} \leftarrow \mathbf{x}_{batch}$ \;
         $\mathbf{y}_{bias} \leftarrow \mathbf{y}$\;
         $\hat{\mathbf{x}} =  \hat{\mathbf{x}} + \mathbf{y}_{bias}$\;
          \textbf{Unsupervised} \\
            sample \(\mathbf{h} \sim p(\mathbf{h}|\mathbf{y}^{t},\mathbf{x}^{t})\) by (\ref{eqhp})(\ref{hidden}) \;
            calculate $\langle\frac{\partial E}{\partial \theta}\rangle_{\mathbf{h}}$ by (\ref{liklyhood1})-(\ref{liklyhood4})\; 
           \For {iteration $\leq$ $N_{step}$}
           {
               sample \(\mathbf{h}^{t,n} \sim p(\mathbf{h}|\mathbf{y}^{t,n}, \mathbf{x}^{t,n})\)  by (\ref{eqhp})(\ref{hidden})\; 
               sample \(\mathbf{y}^{t,n} \sim p(\mathbf{y}|\mathbf{x}^{t,n}, \mathbf{h}^{t,n})\)  by (\ref{eqvp})(\ref{visible})\;                
            } 
            calculate $\langle\frac{\partial E}{\partial \theta}\rangle_{\mathbf{h},\mathbf{y}}$ by (\ref{liklyhood1})-(\ref{liklyhood4}) \; 
            update $\{w_{fu}, w_{fs}, w_{fa}, \mathbf{b}_{h},\mathbf{b}_{x}\}$ by (\ref{eqderi}) and (\ref{sparsity})\; 
         $\hat{\mathbf{h}} \leftarrow \mathbf{h}$ \;
         \textbf{Backward} \\
          $ \Delta = \|\hat{\mathbf{h}}-target\|$ \;
          $\frac{\partial \Delta}{\partial J} $, $\frac{\partial \Delta}{\partial W} $, $\frac{\partial \Delta}{\partial M} $, $\frac{\partial \Delta}{\partial F^{(i)}} $\;
           update $\{F^{(i)}, M^{m}, J, W_{fx}, W_{fy}, W_{fh}\}$ \;            
    }
  } 
  \KwOut{parameter set} 
\end{algorithm}

For NES-I and NES-B models, the training procedure is similar to the algorithm 1, and the difference lies that both models are updated based on a standard RBM pre-training without introducing factored matrix parameters.

\section{Experiments}
In this study, we conduct the experimental evaluation of the proposed three NES models on a public dataset KARA ONE \cite{zhao2015classifying}. This dataset combines 3 modalities (EEG, face tracking, and audio) during imagined and vocalized phonemic and single-word prompts. In their EEG recording experiments, each participant was instructed to look at the computer monitor and move as little as possible. Seven phonemic/syllabic prompts and 4 words were used in several repeated experimental trials, and each one of 14 participants produced 132 trials. Each trial consisted of 4 successive states, including a rest state, a stimulus state, an imagined speech state, and a speaking state. The EEG data were collected using a 64-channel Neuroscan Quick-cap, and the electrode placement followed the 10-20 system. In this dataset, 62 channel EEG data ( M1 and M2 channels are excluded as suggested by the original dataset) are used for evaluations of the proposed NES models. 

\subsection{Experimental Data Process}
A general data pre-process is applied to the EEG data, and the functions including removal of ocular artifacts using blind source separation are modified from EEGLAB (i.e., runica) \cite{zhao2015classifying}. The EEG data are band-pass filtered between 1 Hz and 200 Hz (i.e., the butterworth filter), and the mean values are subtracted from each channel data. The EEG data are segmented into different trials, and each trial is further segmented into the 4 states described in above section, among which the EEG signals associated with imagined speech and spoken speech are used as the feed. All EEG data are normalized in each channel, and the batch data of speech are also normalized in time domain. In each channel, the window size is fixed as 20 ms and a hop size 10 ms (i.e., 100 segments in each second). The speech signals are downsampled into 1kHz to match the sampling rate of EEG signals. After the pre-filtering, the envelope of speech power is extracted as the training data. To implement the batch data processing, speech signals and EEG signals are padded (i.e., the shorter signals will be extended by filling with zeros).
\begin{figure*}[hbt]
\center
\vspace{-5mm}
\subfigure{\includegraphics[scale=0.24]{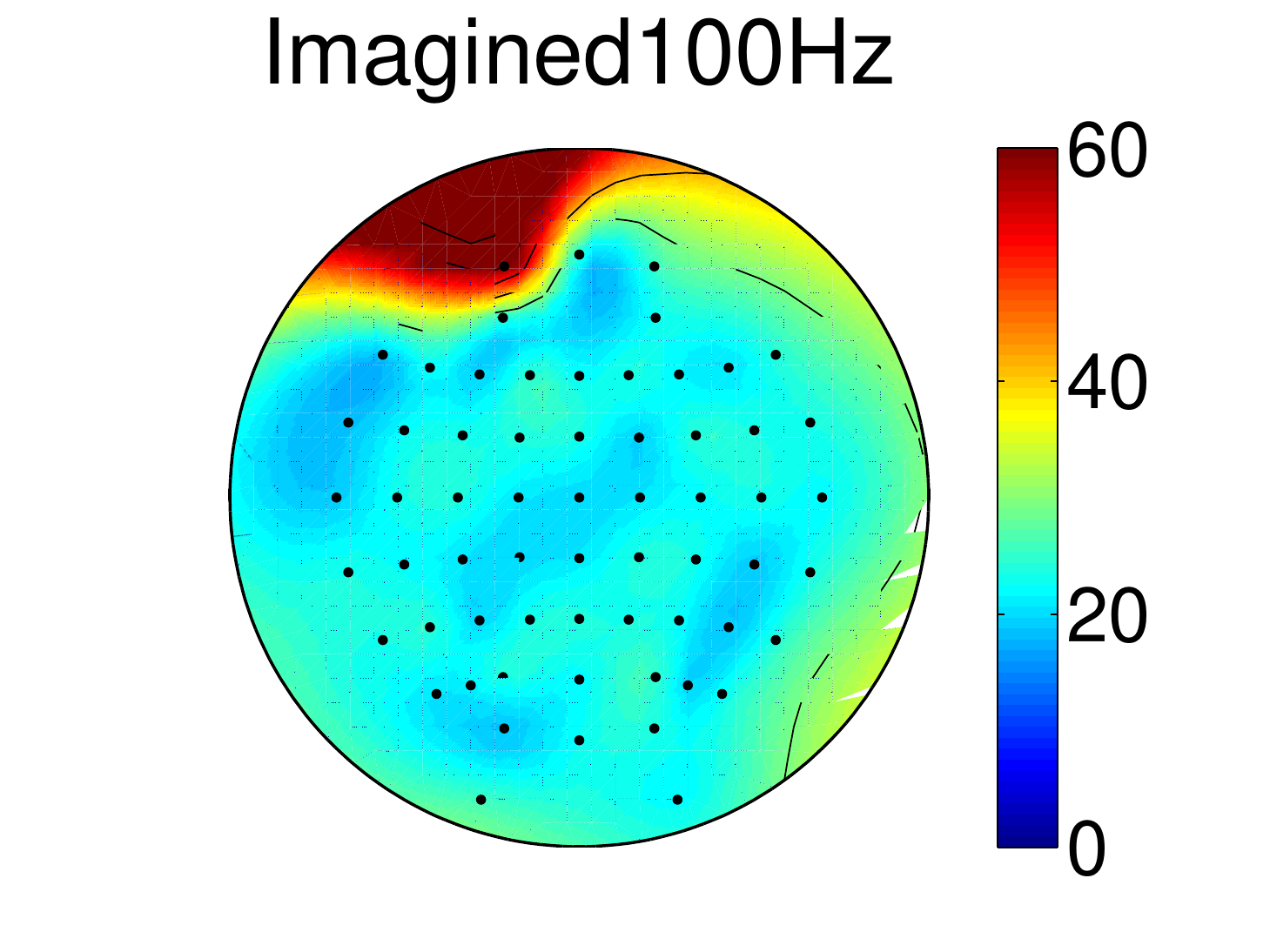}}
\hspace{-0.8em}
\subfigure{\includegraphics[scale=0.24]{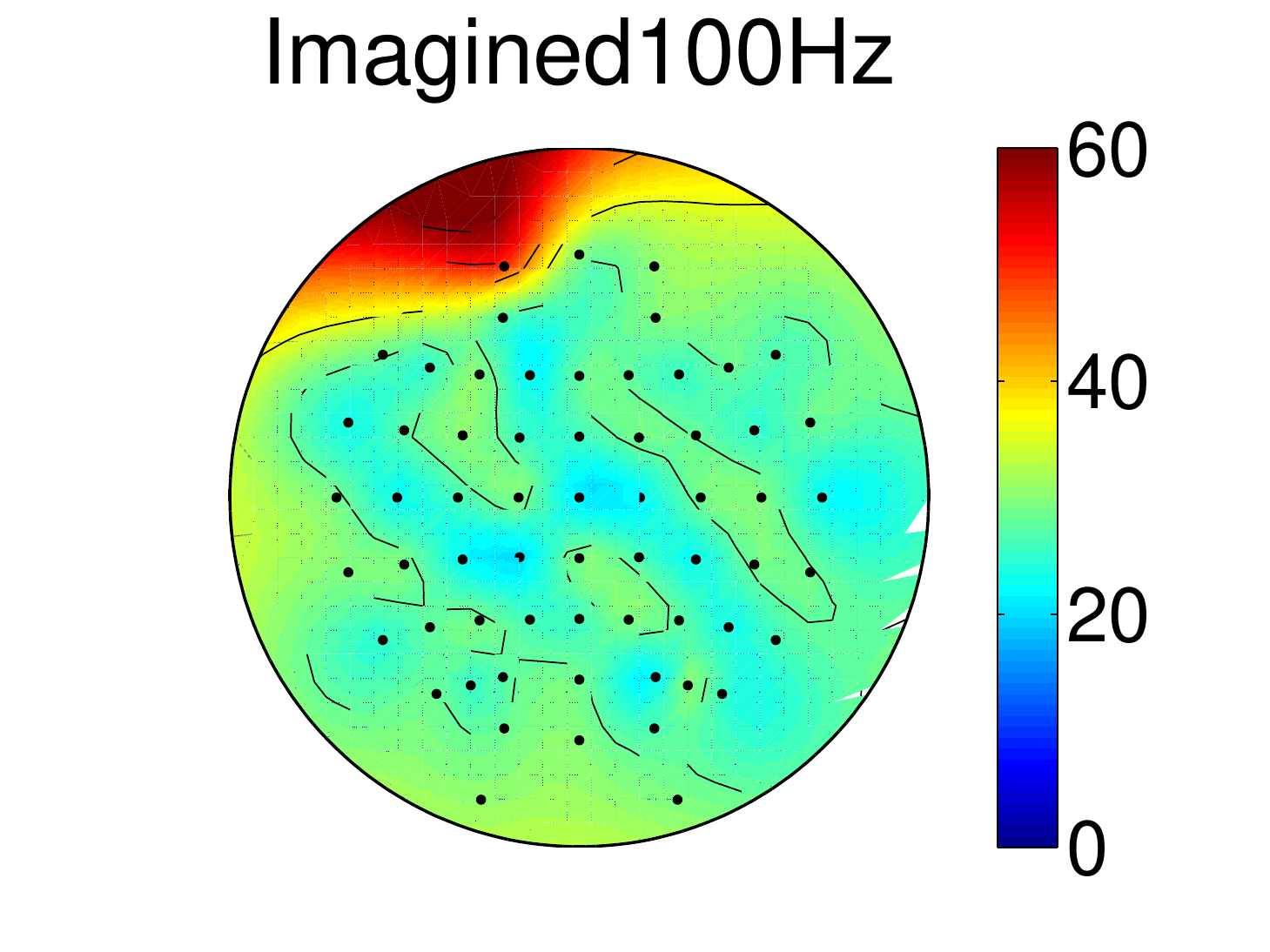}}
\hspace{-0.8em}
\subfigure{\includegraphics[scale=0.24]{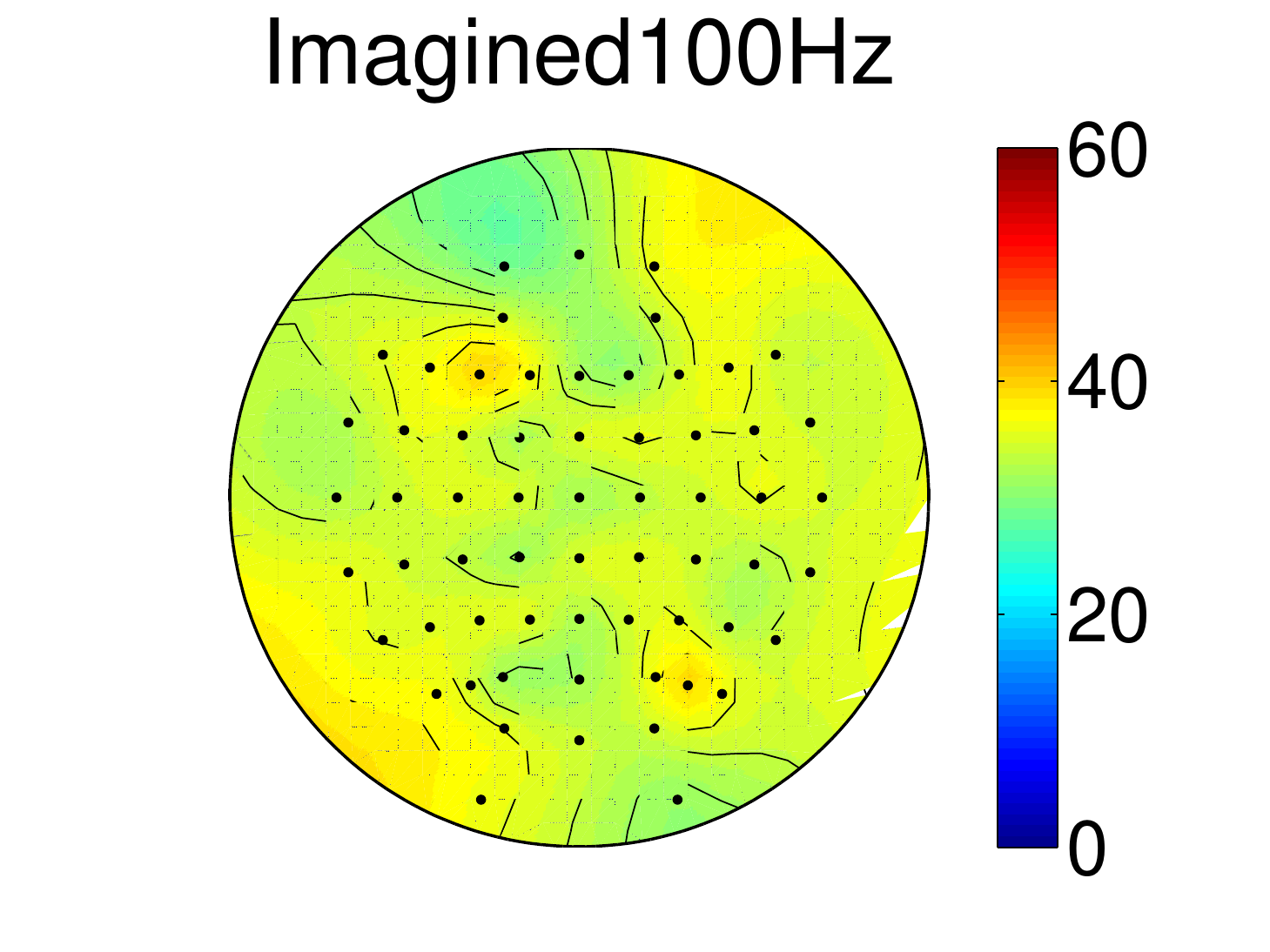}}
\hspace{-0.8em}
\subfigure{\includegraphics[scale=0.24]{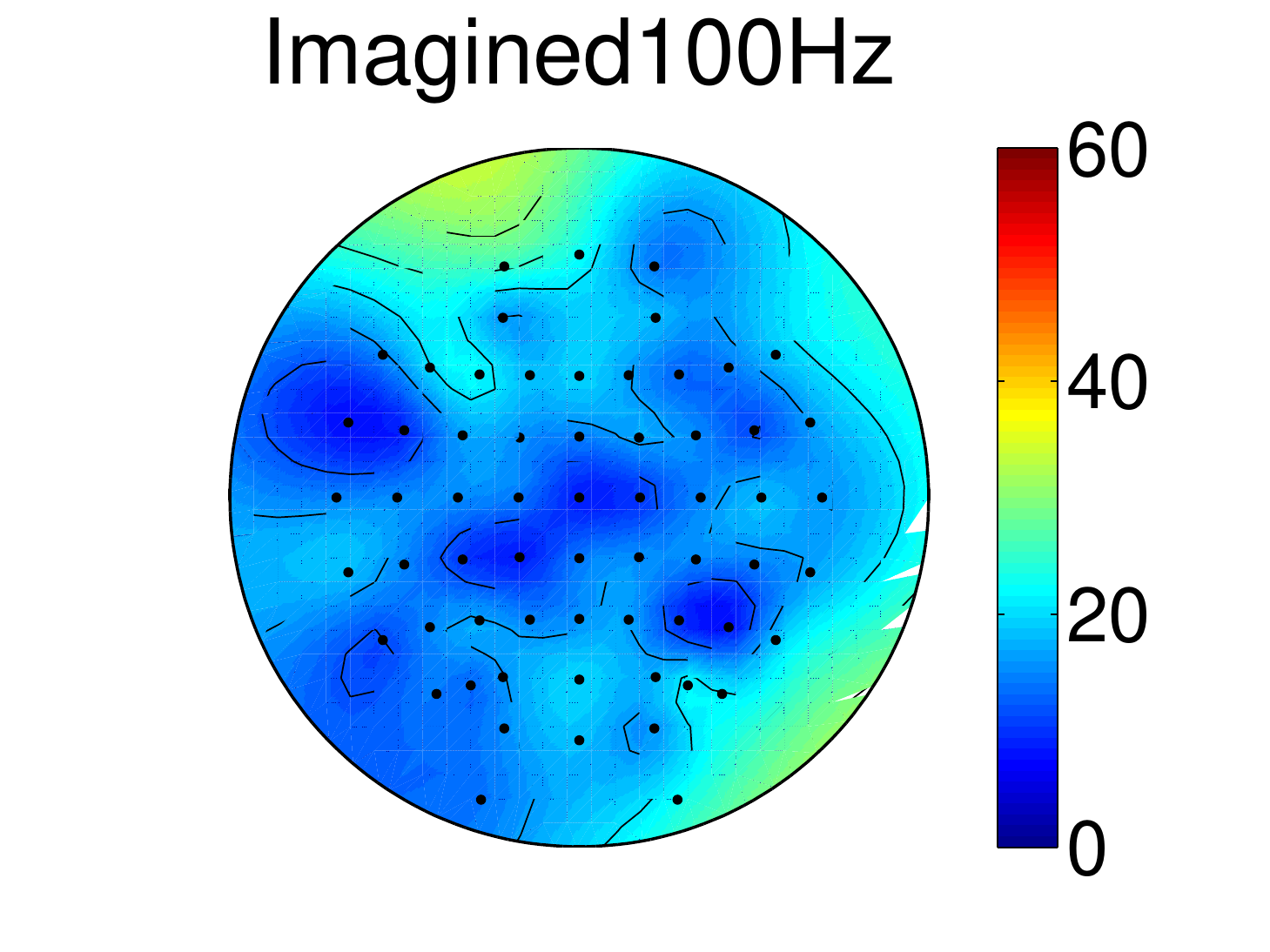}}
\hspace{-0.8em}
\subfigure{\includegraphics[scale=0.24]{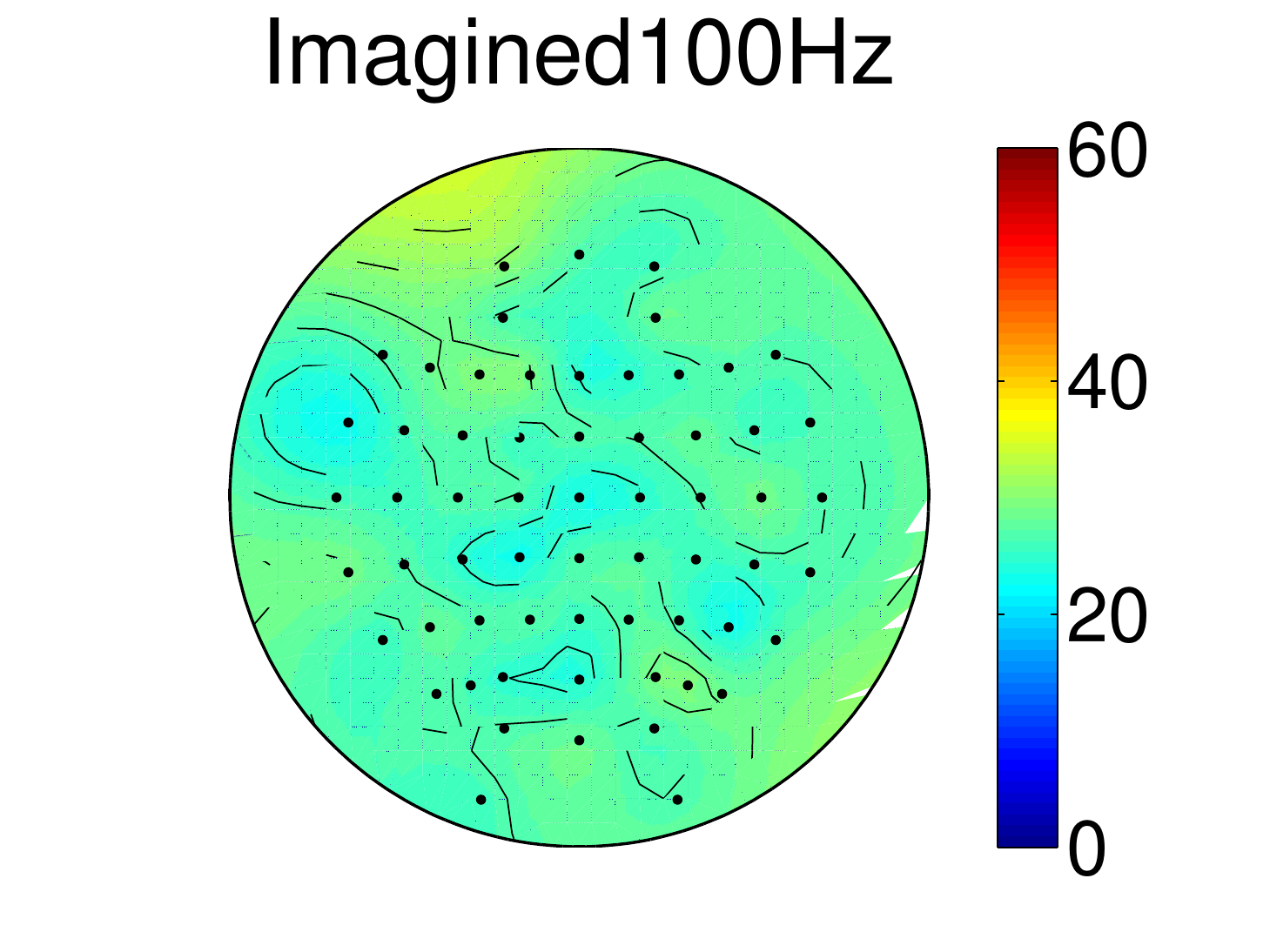}}
\hspace{-0.8em}
\subfigure{\includegraphics[scale=0.24]{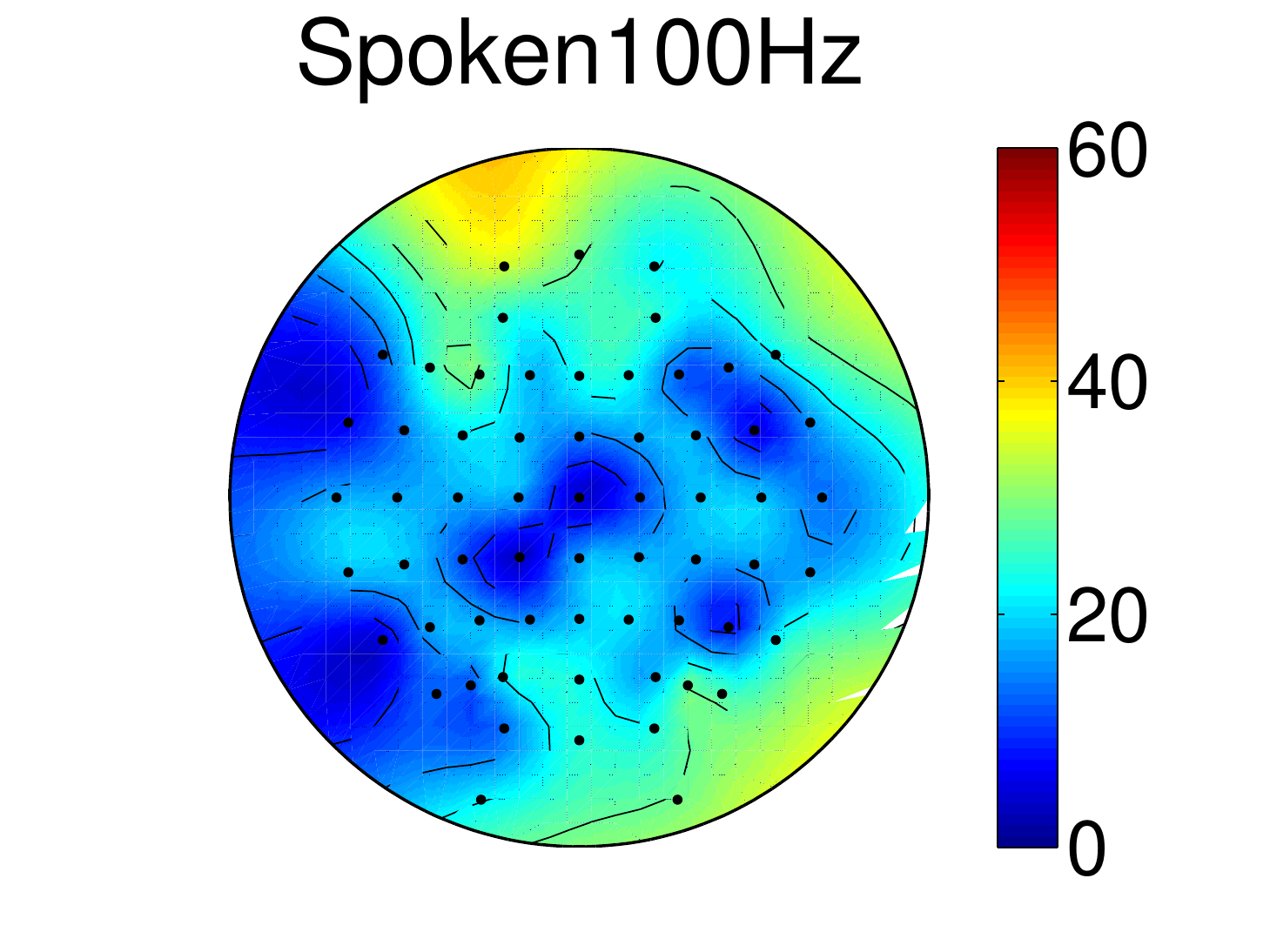}}
\hspace{-0.6em}
\subfigure{\includegraphics[scale=0.24]{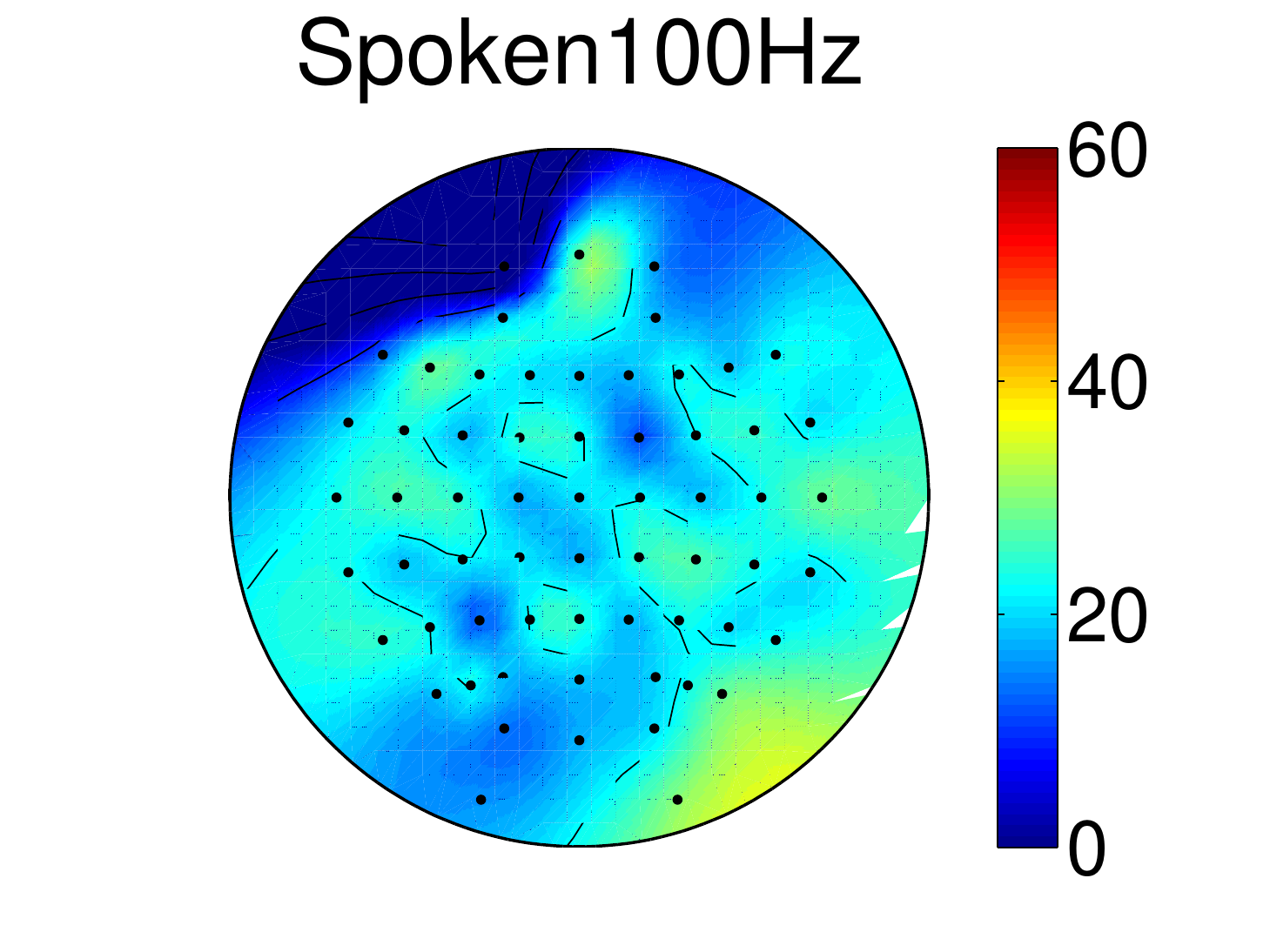}}
\hspace{-0.6em}
\subfigure{\includegraphics[scale=0.24]{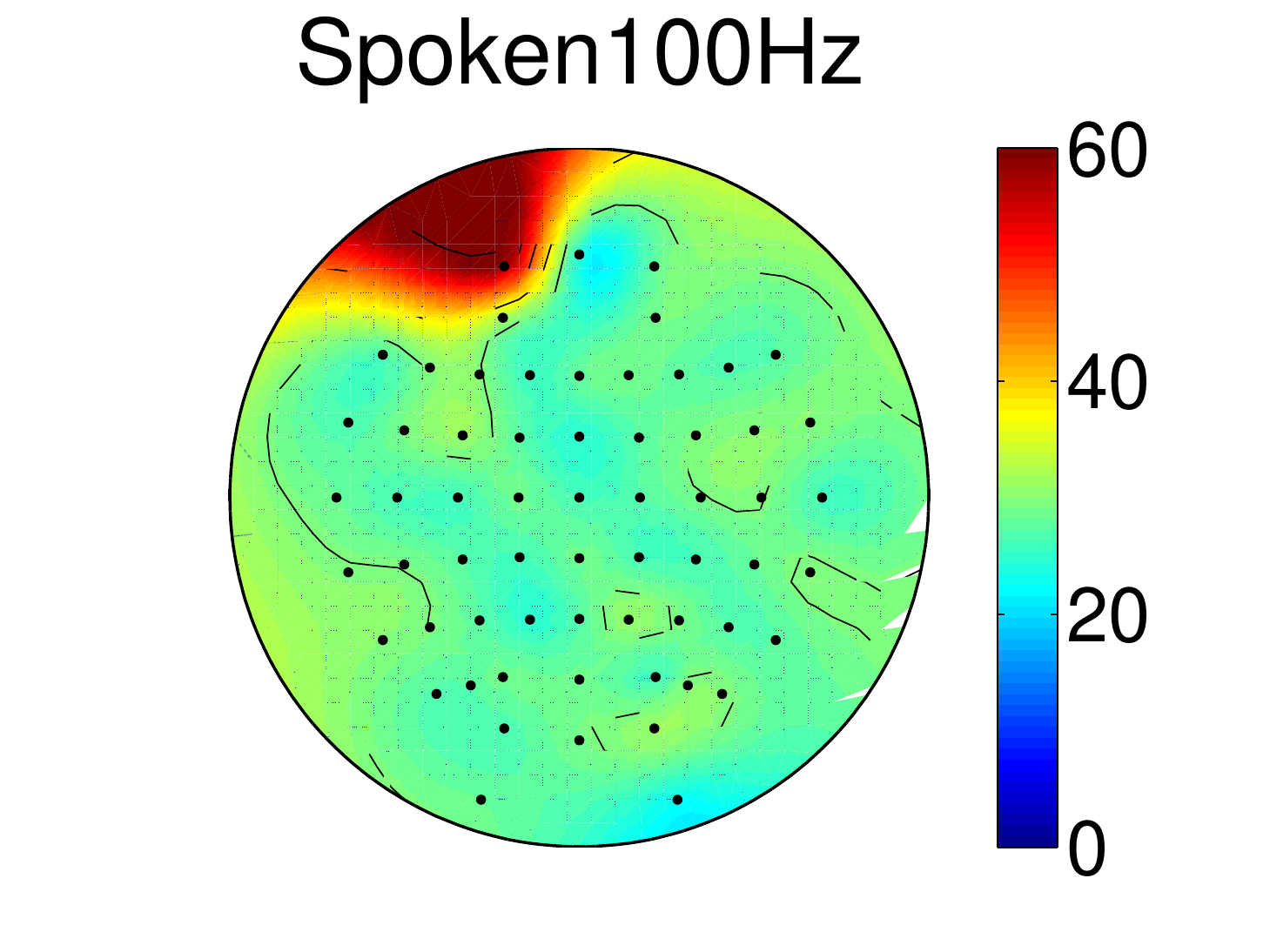}}
\hspace{-0.6em}
\subfigure{\includegraphics[scale=0.24]{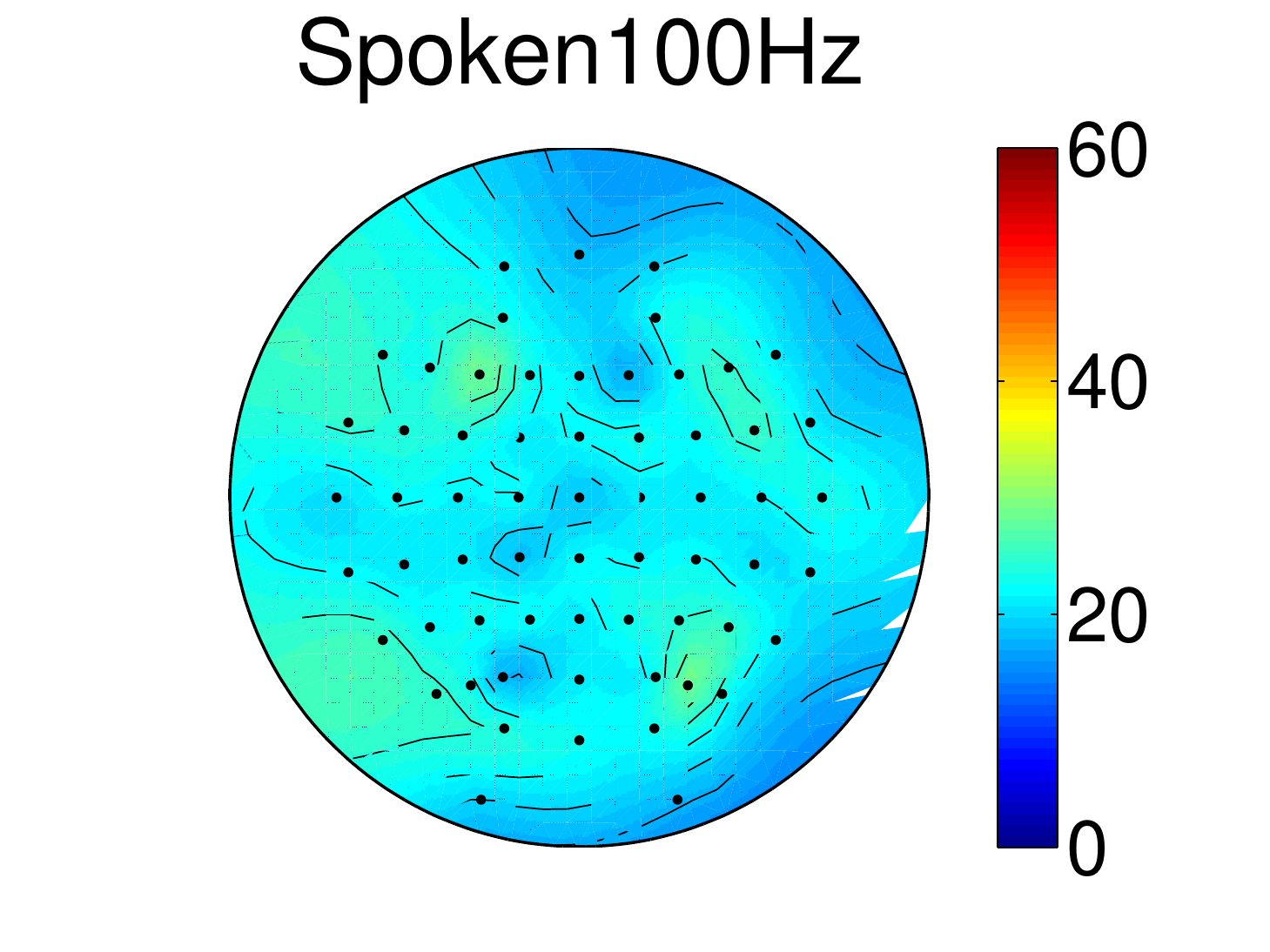}}
\hspace{-0.6em}
\subfigure{\includegraphics[scale=0.24]{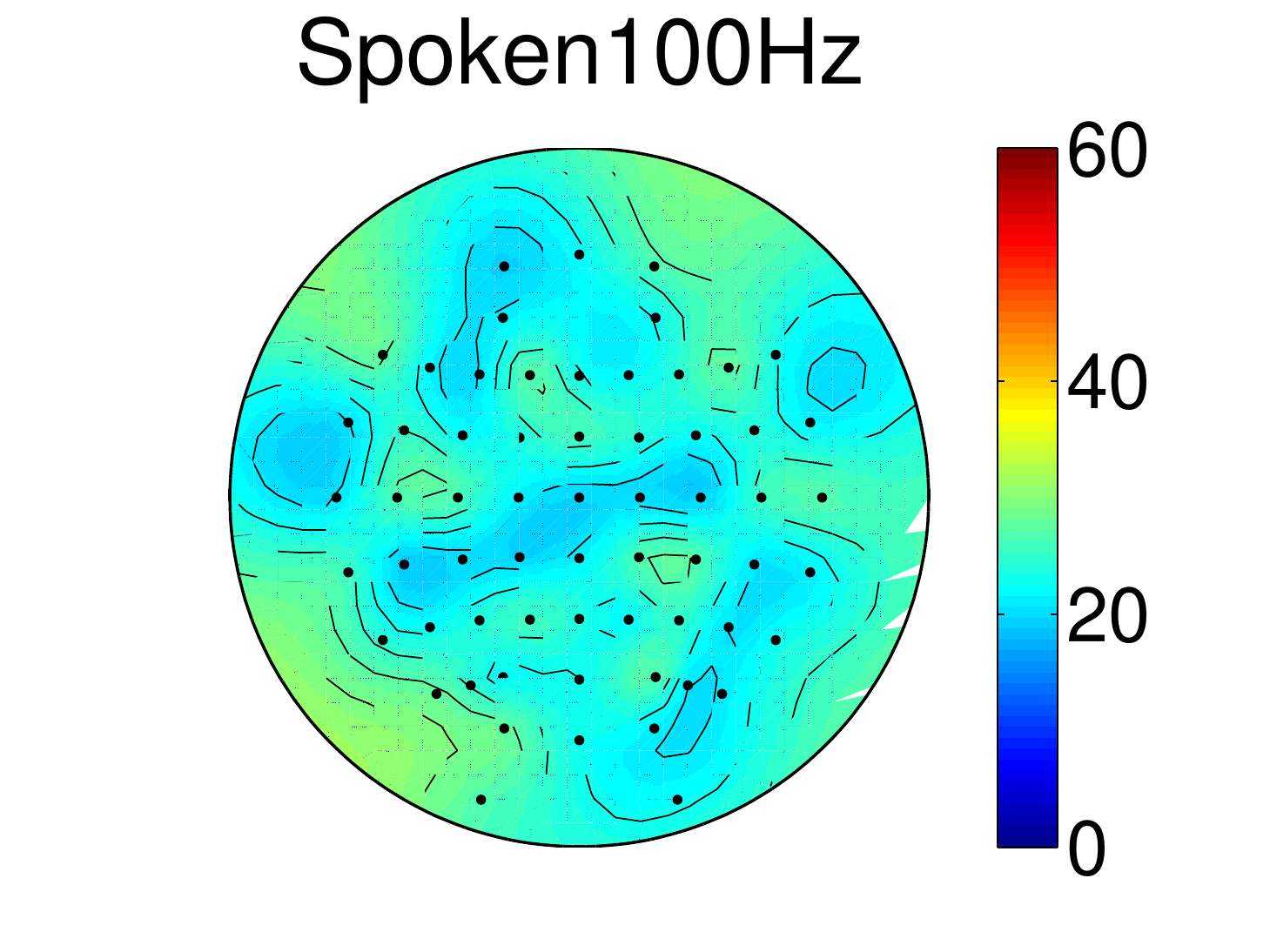}}
\hspace{-0.8em}
\subfigure{\includegraphics[scale=0.24]{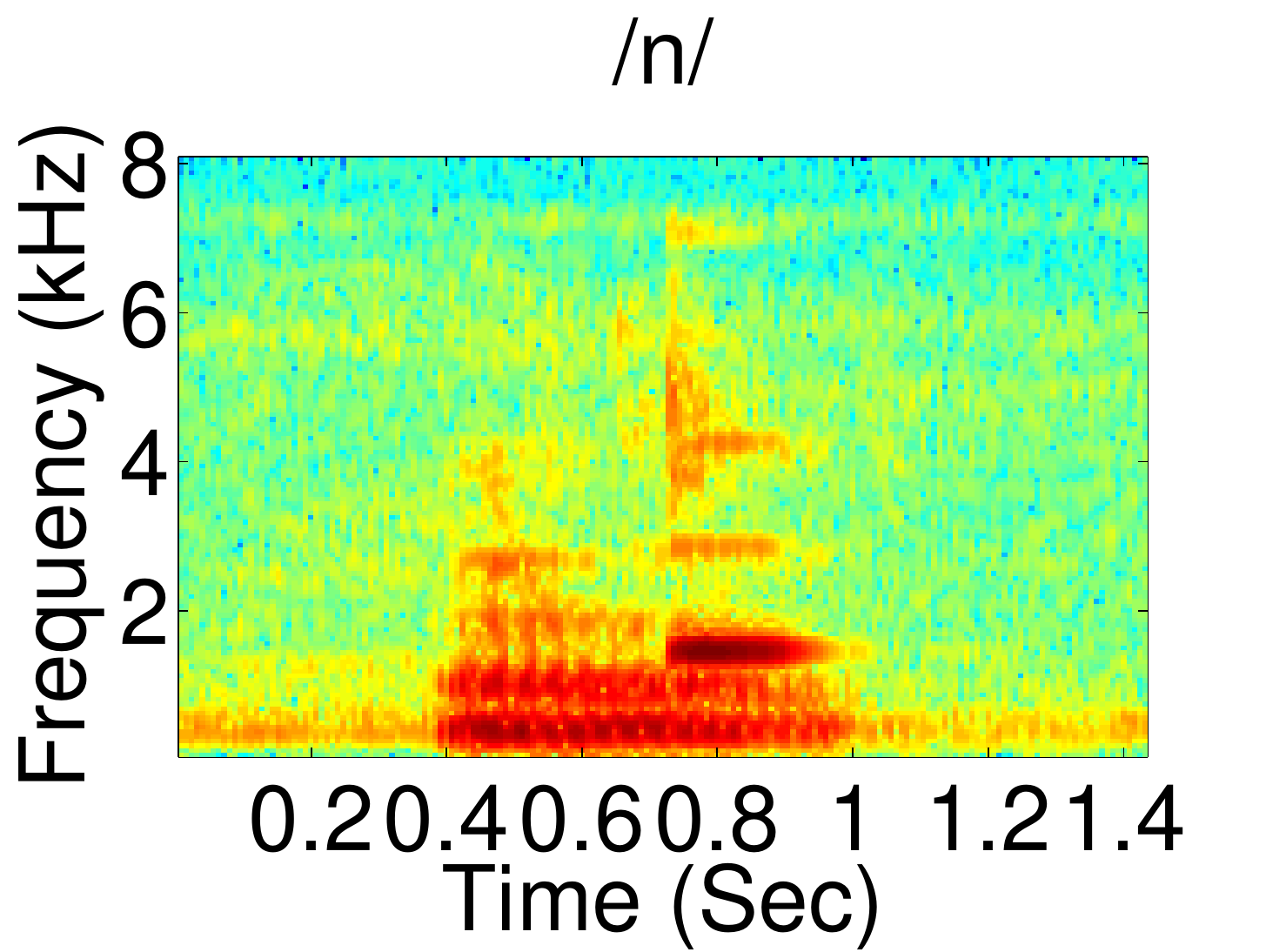}}
\hspace{-0.8em}
\subfigure{\includegraphics[scale=0.24]{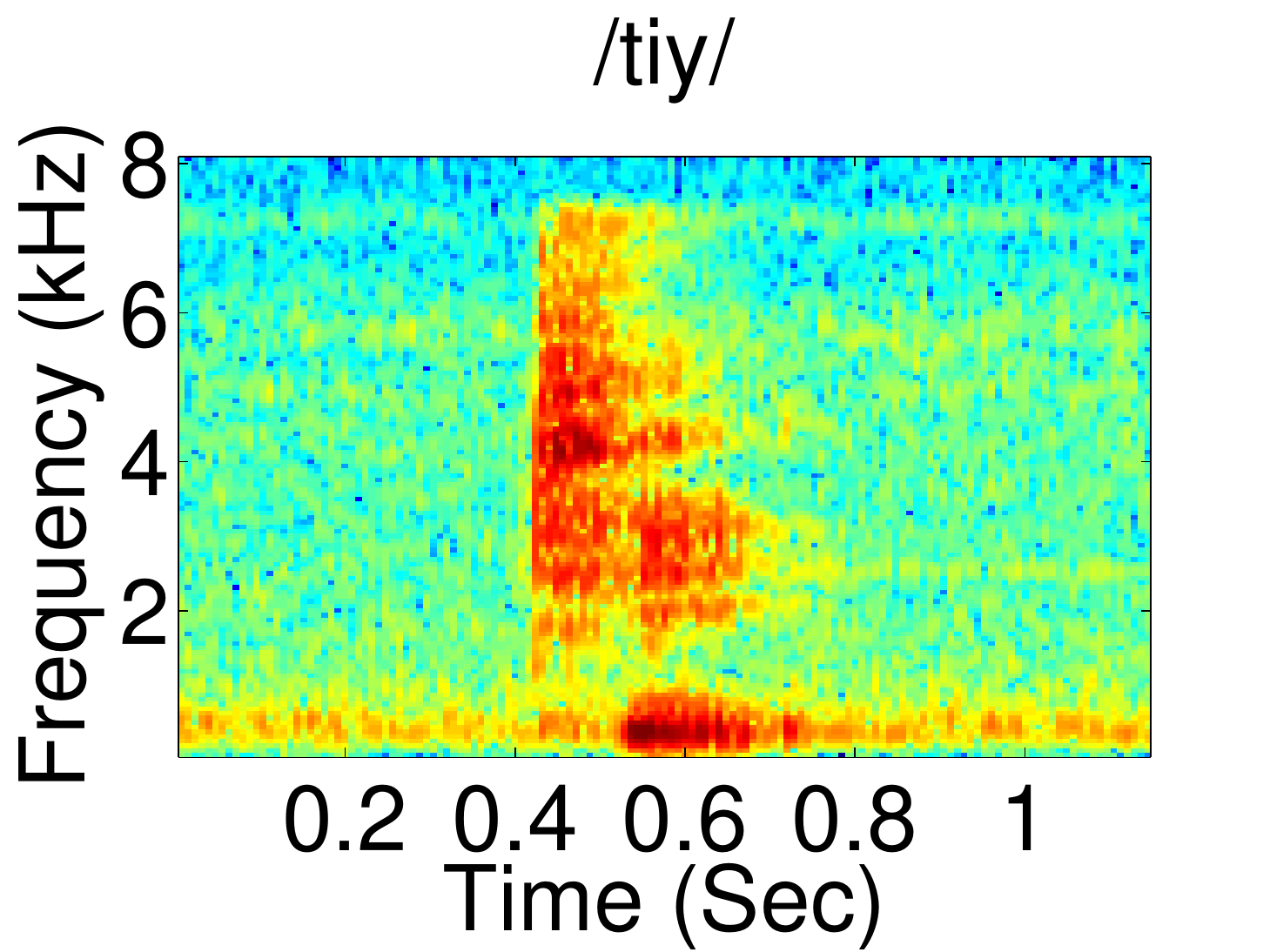}}
\hspace{-0.8em}
\subfigure{\includegraphics[scale=0.24]{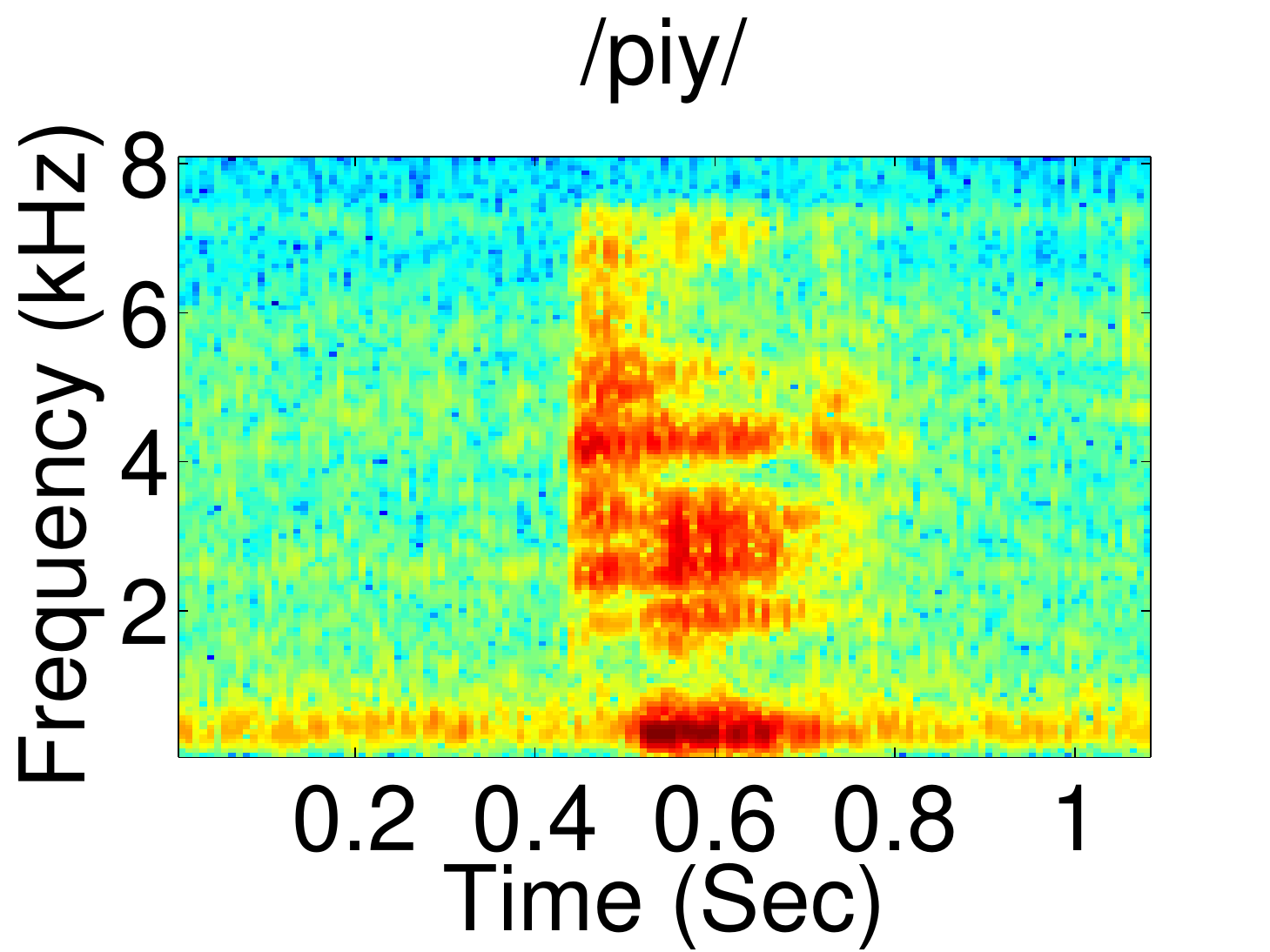}}
\hspace{-0.8em}
\subfigure{\includegraphics[scale=0.24]{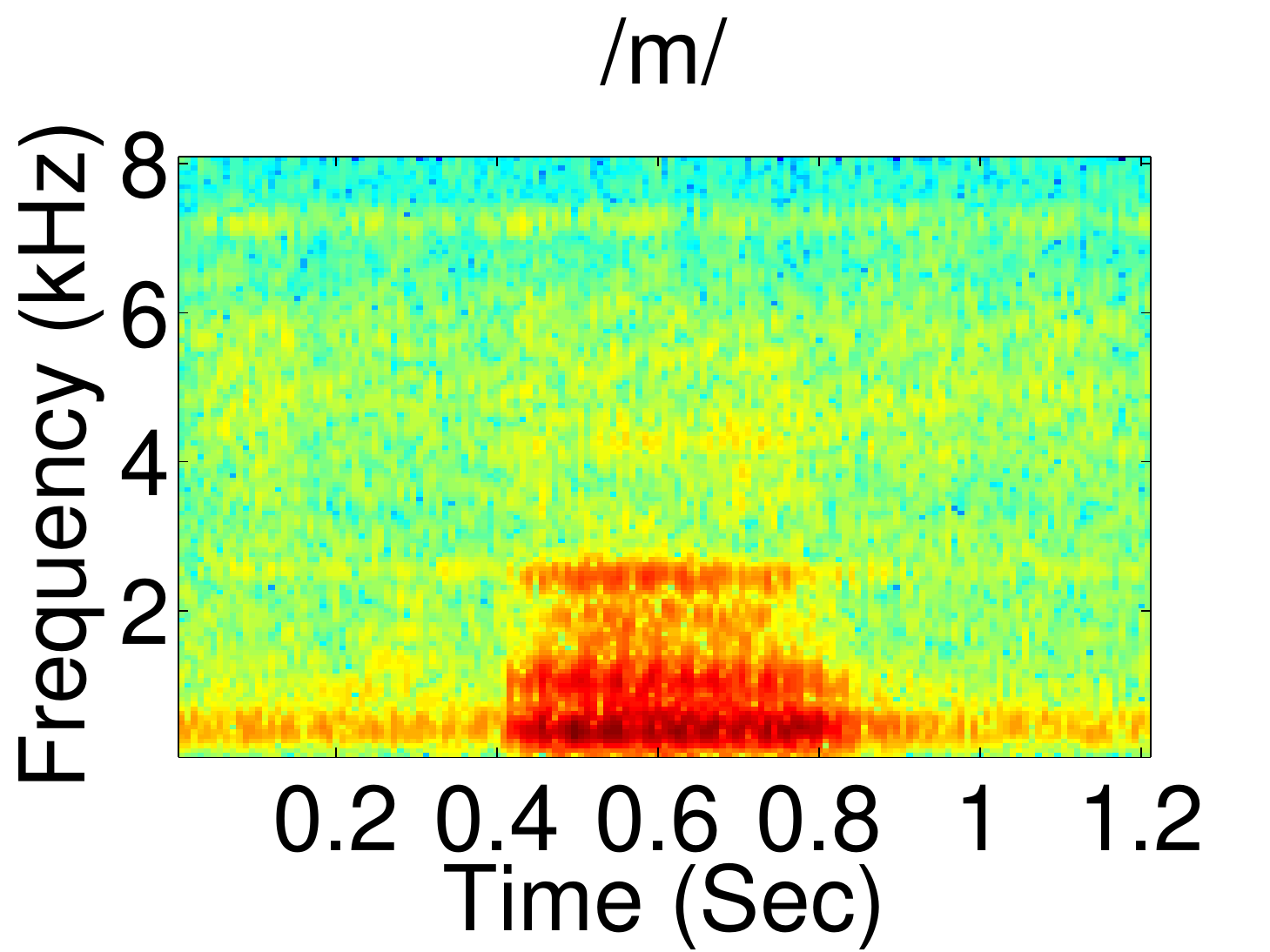}}
\hspace{-0.8em}
\subfigure{\includegraphics[scale=0.24]{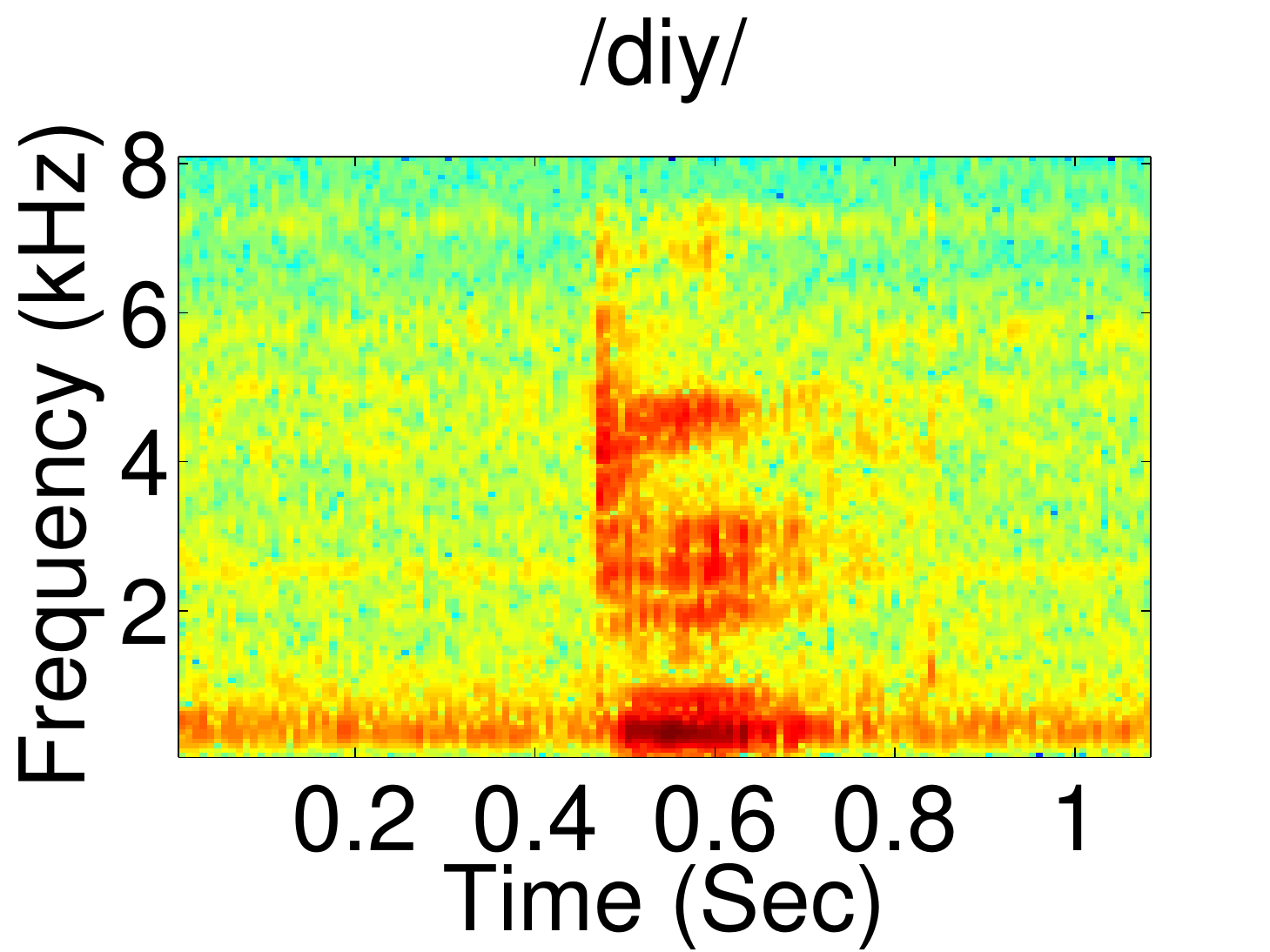}}
\hspace{-1em}
\subfigure{\includegraphics[scale=0.24]{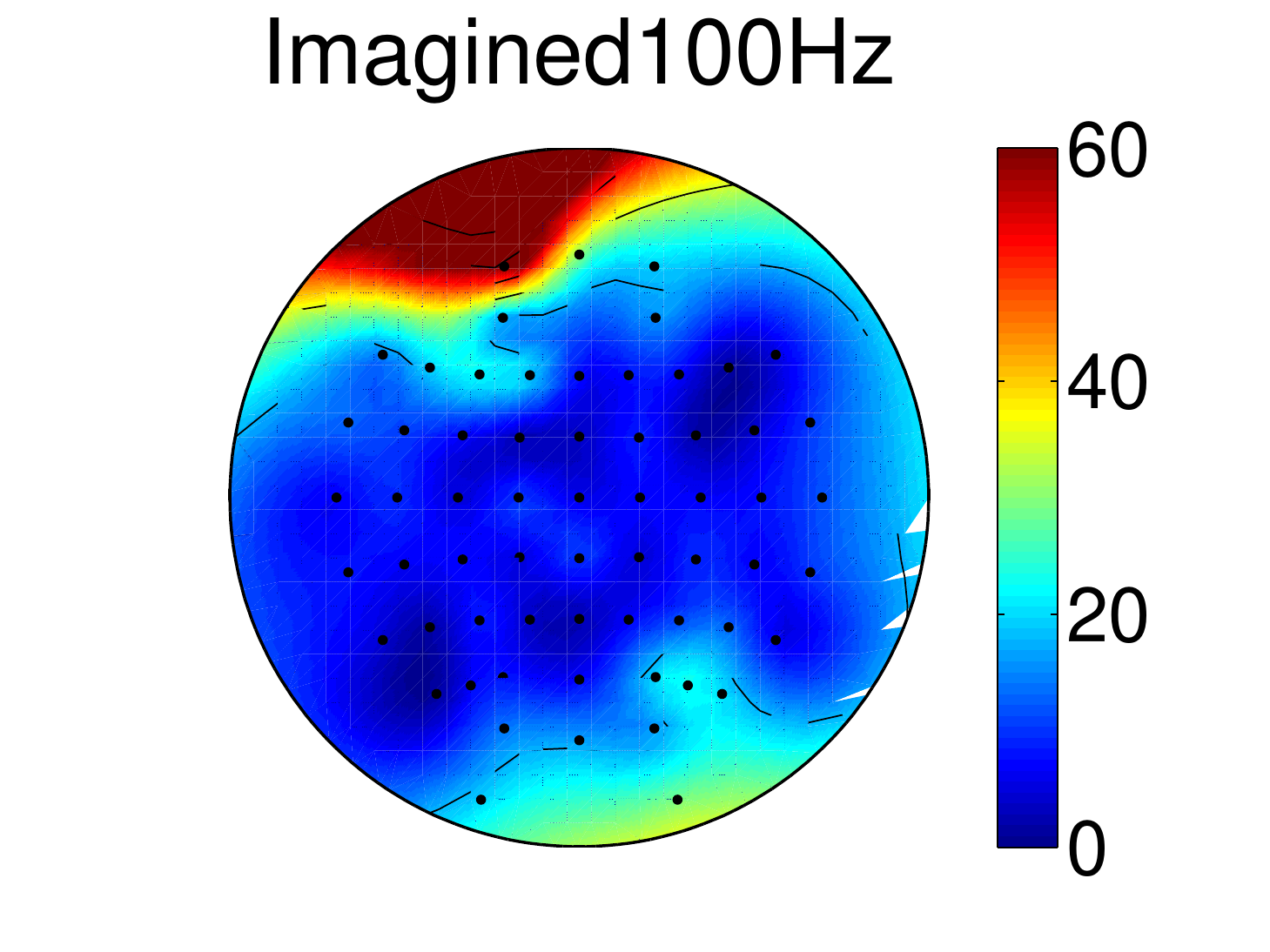}}
\hspace{-0.6em}
\subfigure{\includegraphics[scale=0.24]{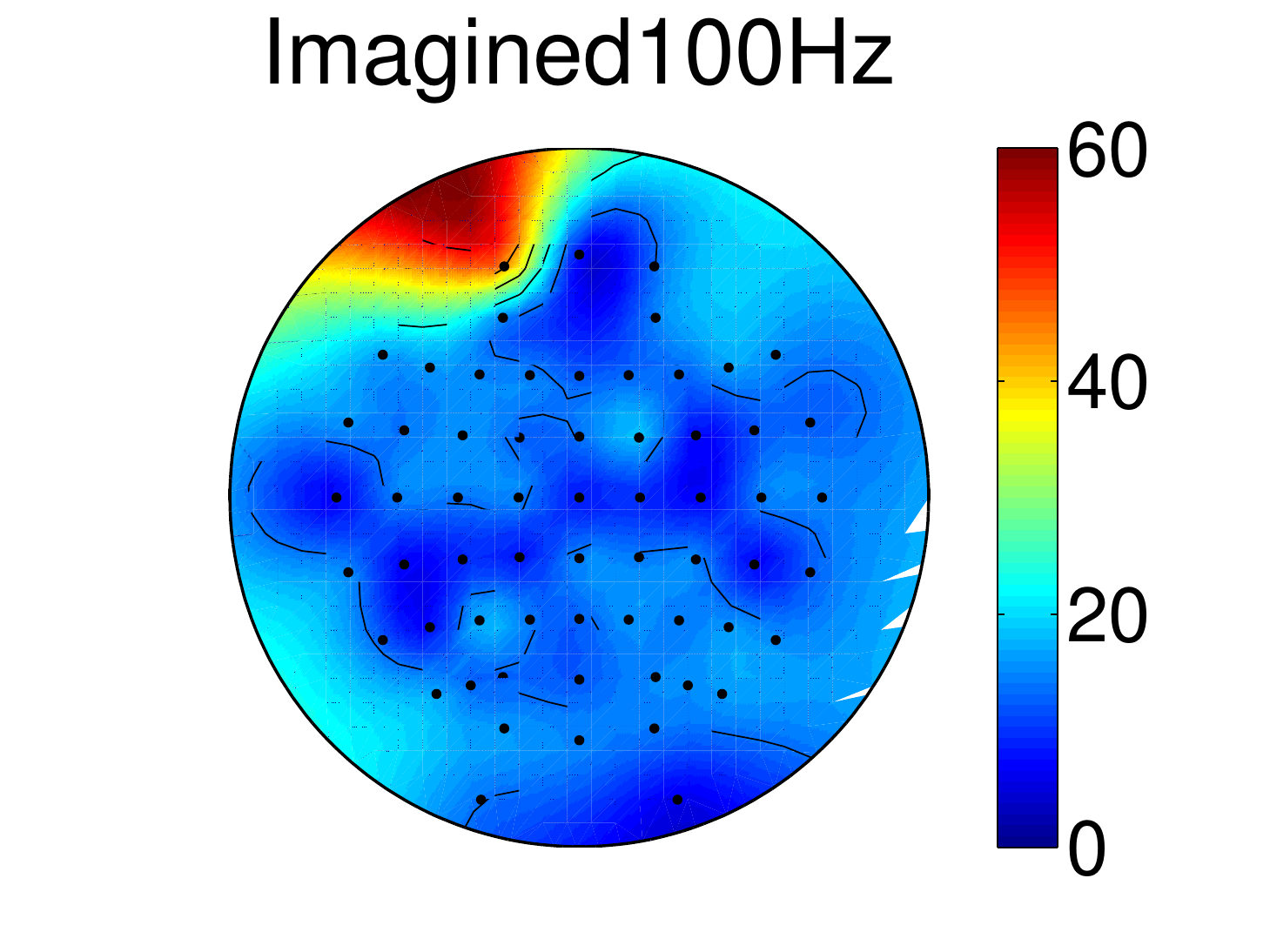}}
\hspace{-0.6em}
\subfigure{\includegraphics[scale=0.24]{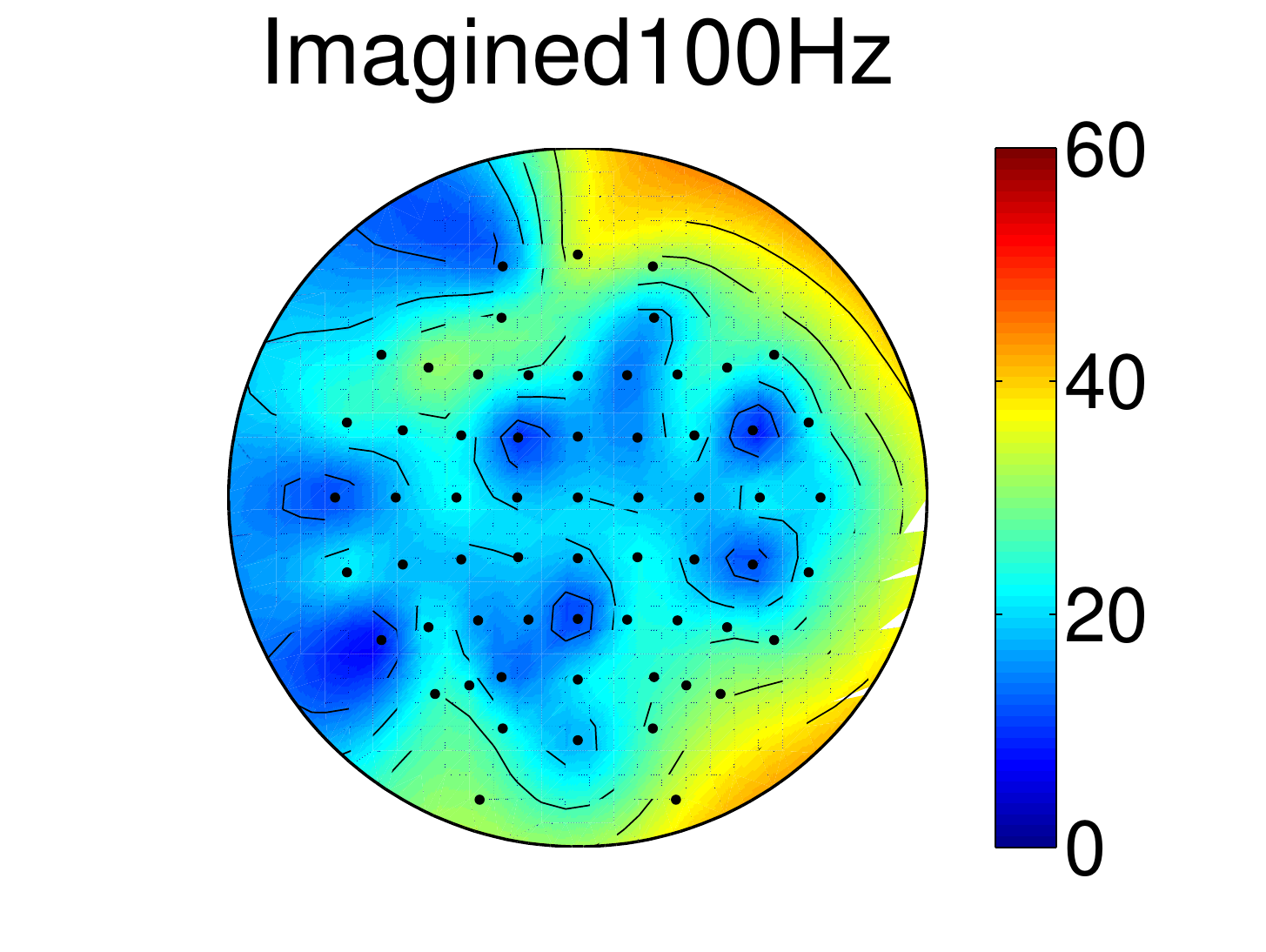}}
\hspace{-0.6em}
\subfigure{\includegraphics[scale=0.24]{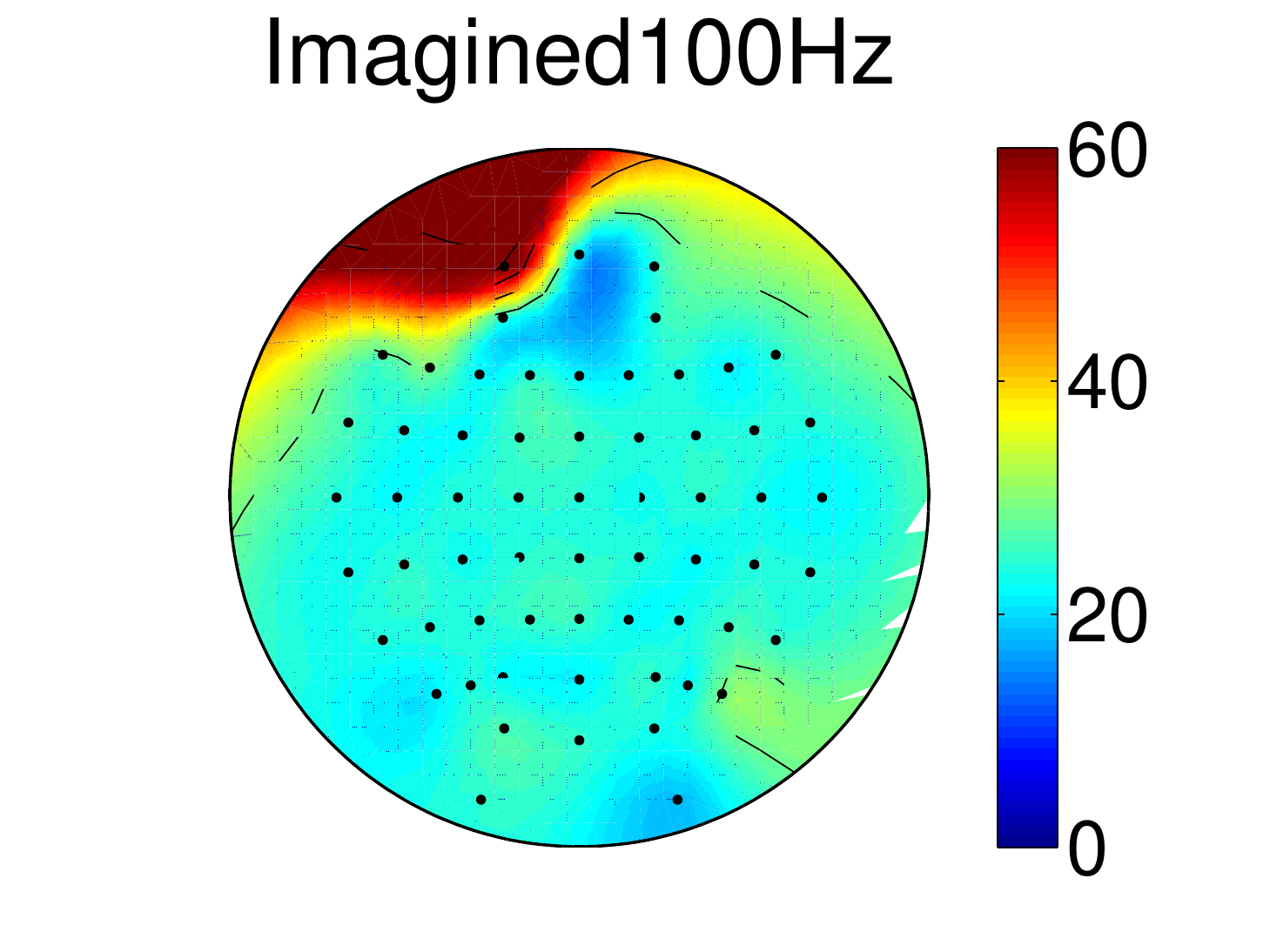}}
\hspace{-0.6em}
\subfigure{\includegraphics[scale=0.24]{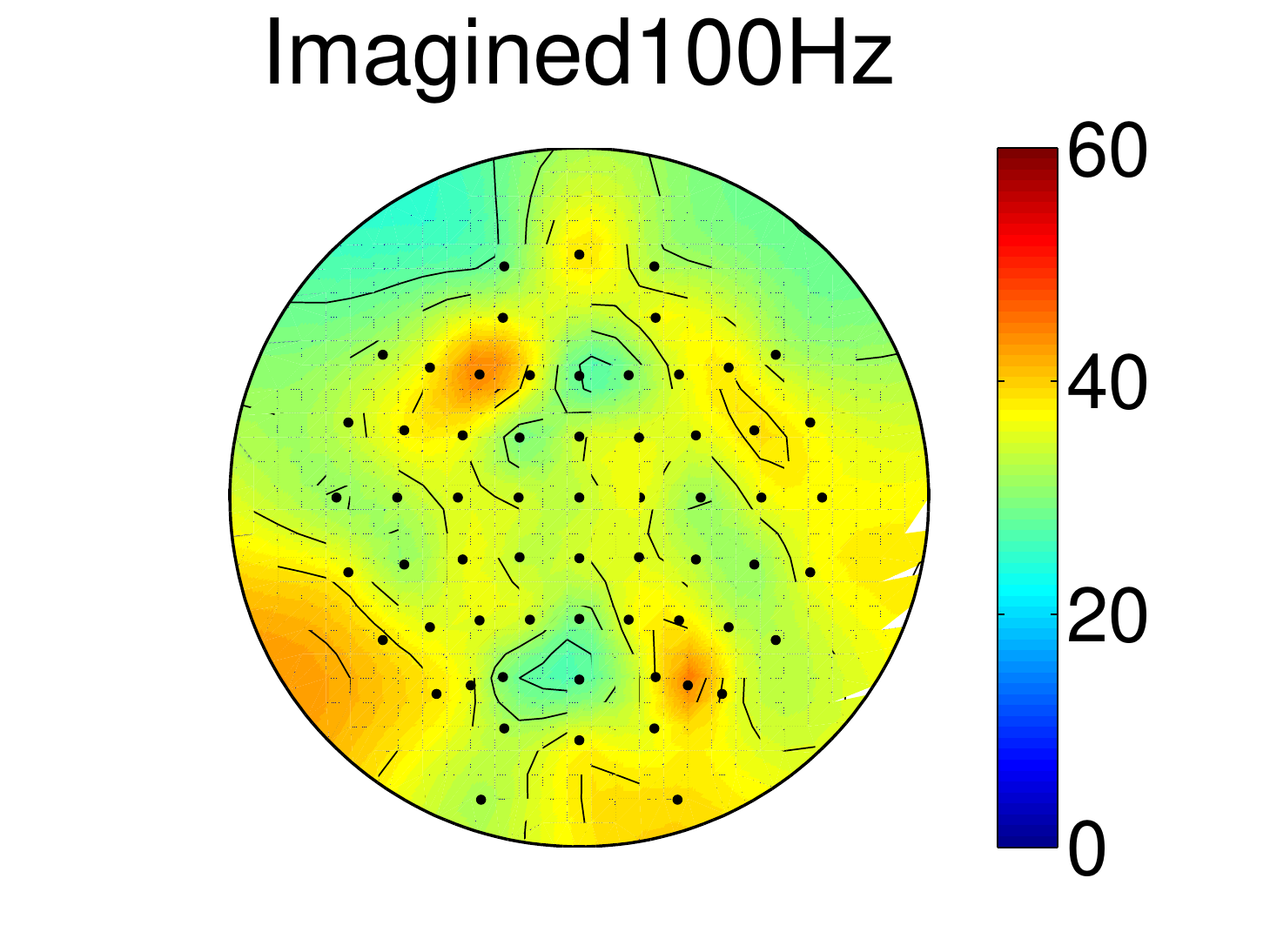}}
\hspace{-0.8em}
\subfigure{\includegraphics[scale=0.24]{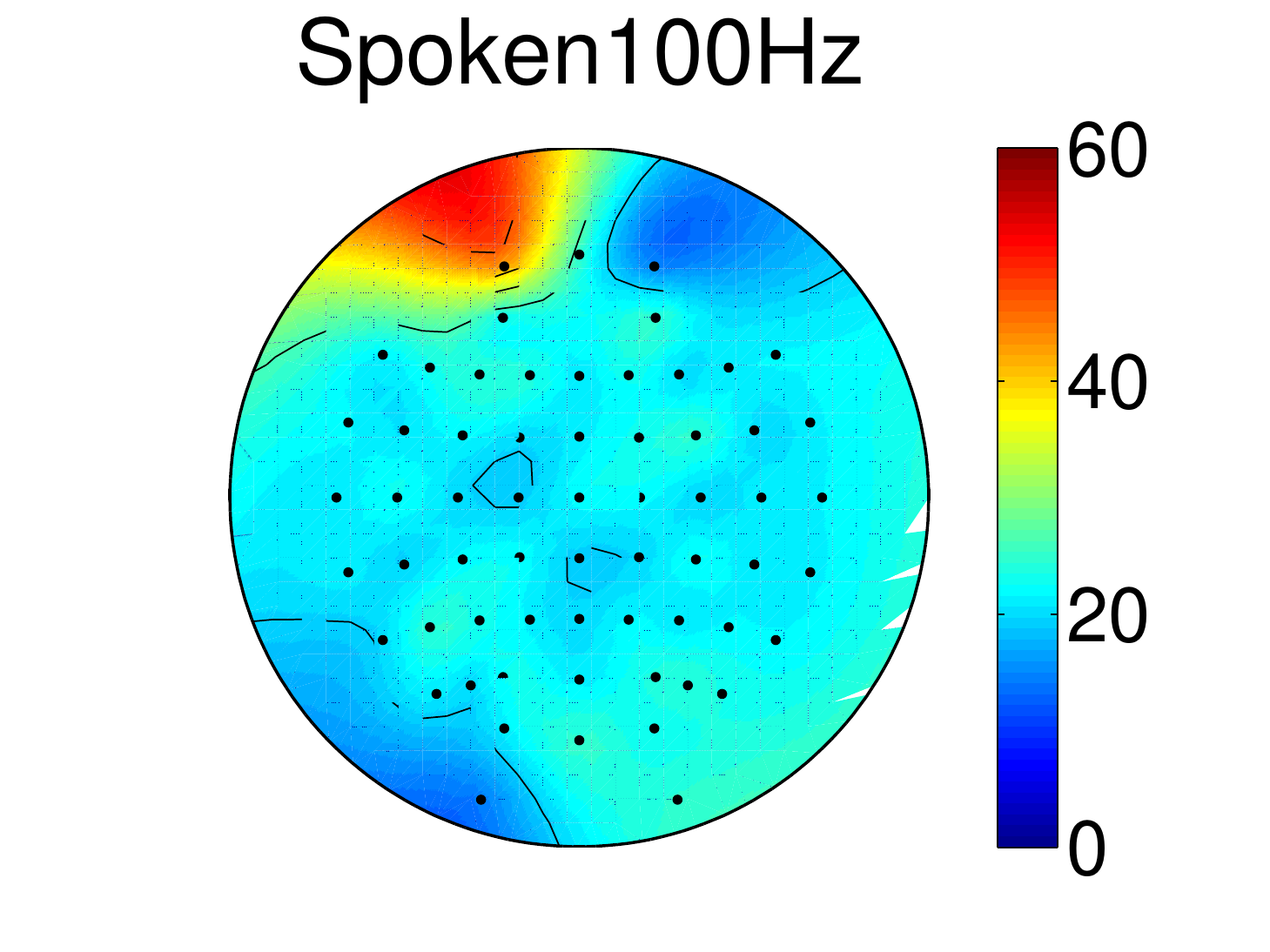}}
\hspace{-0.8em}
\subfigure{\includegraphics[scale=0.24]{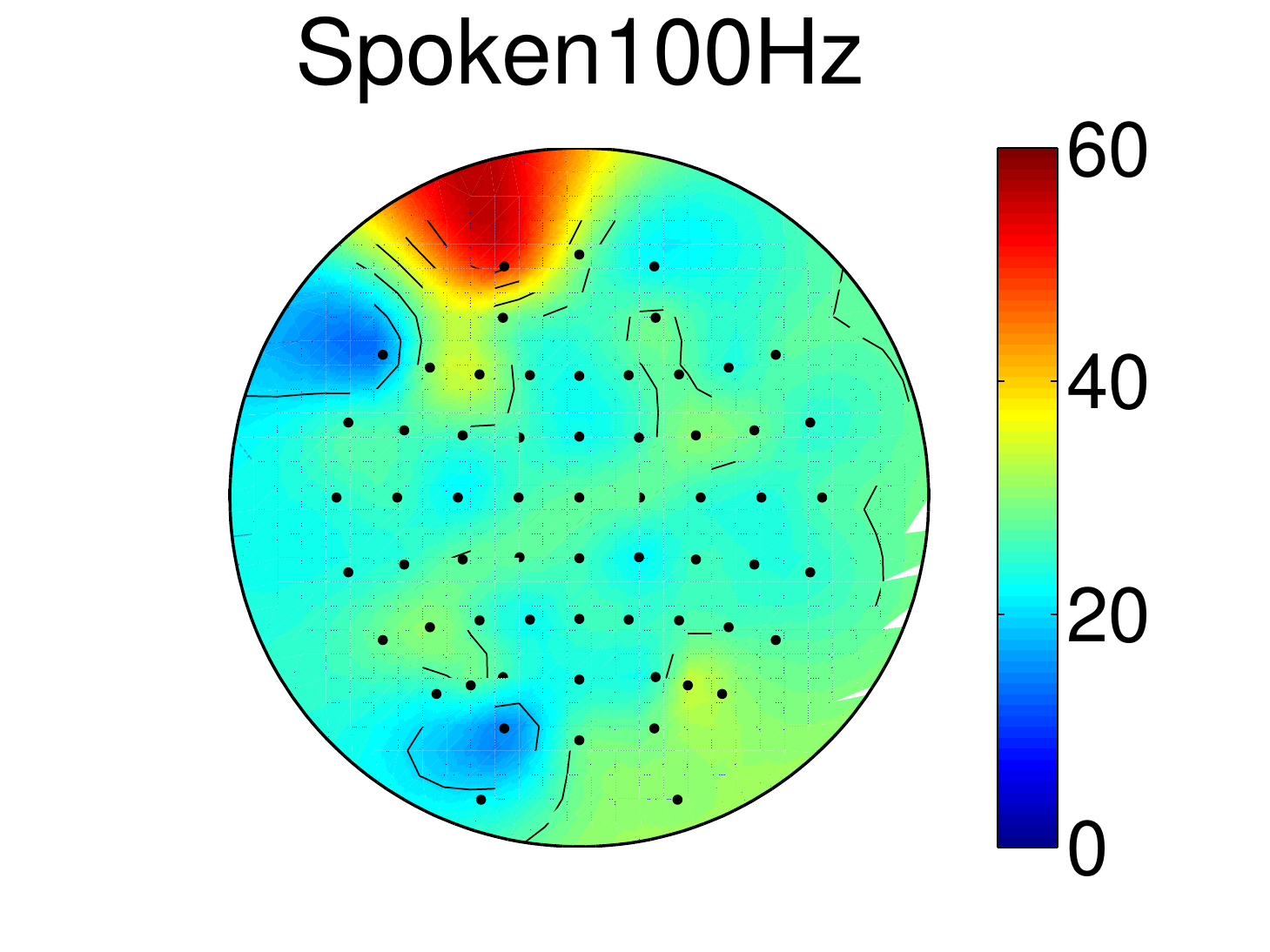}}
\hspace{-0.6em}
\subfigure{\includegraphics[scale=0.24]{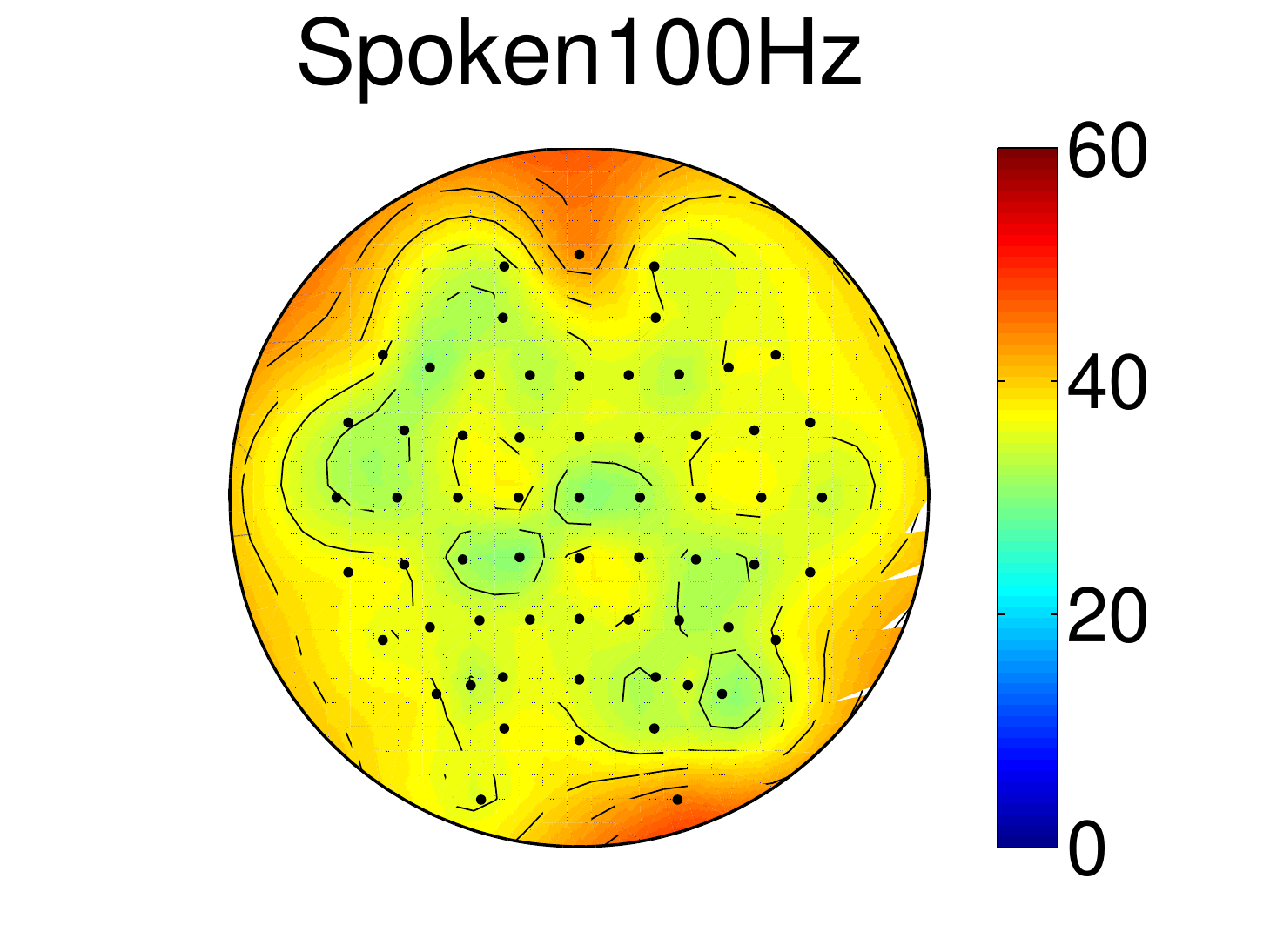}}
\hspace{-0.6em}
\subfigure{\includegraphics[scale=0.24]{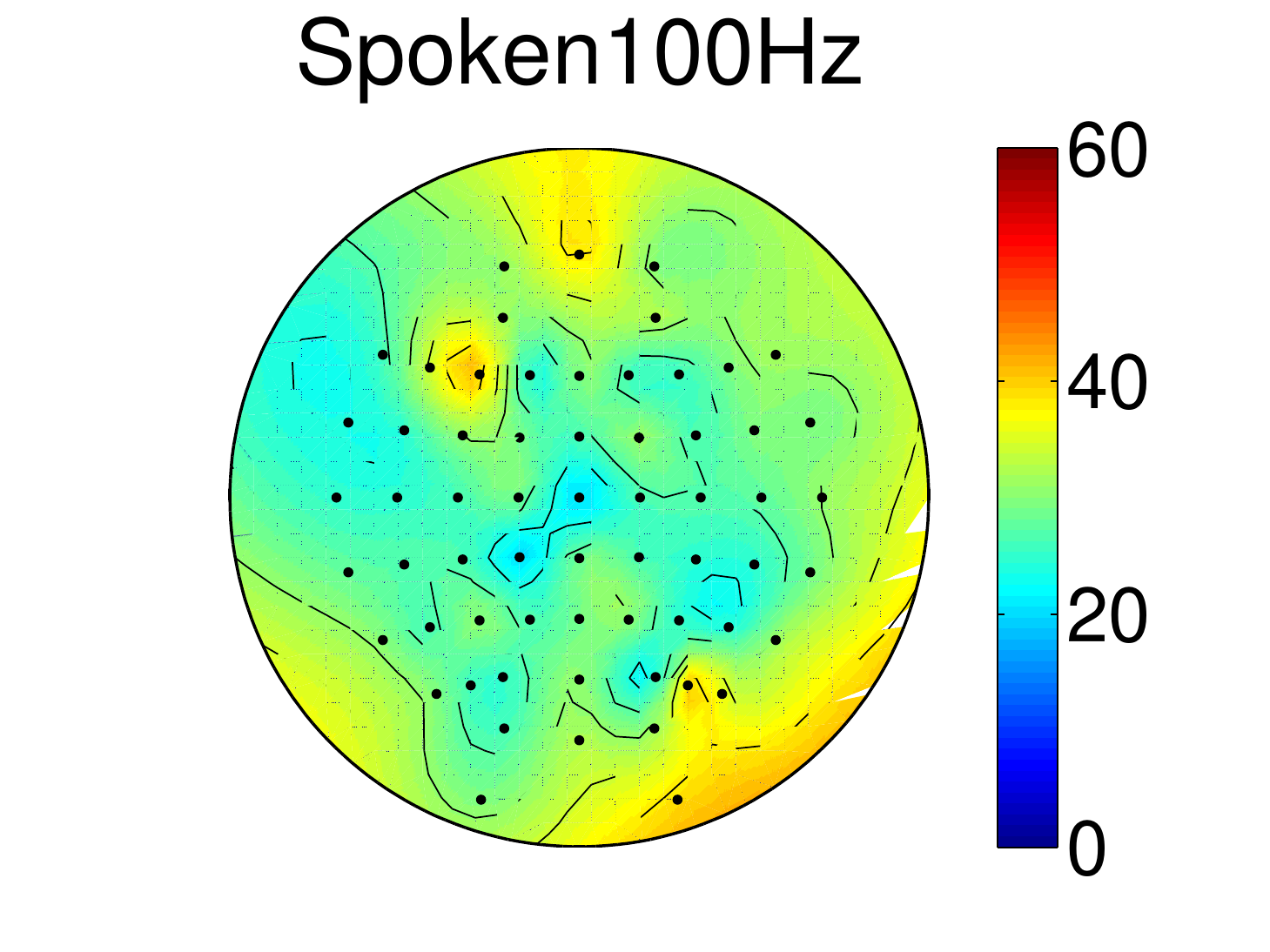}}
\hspace{-0.6em}
\subfigure{\includegraphics[scale=0.24]{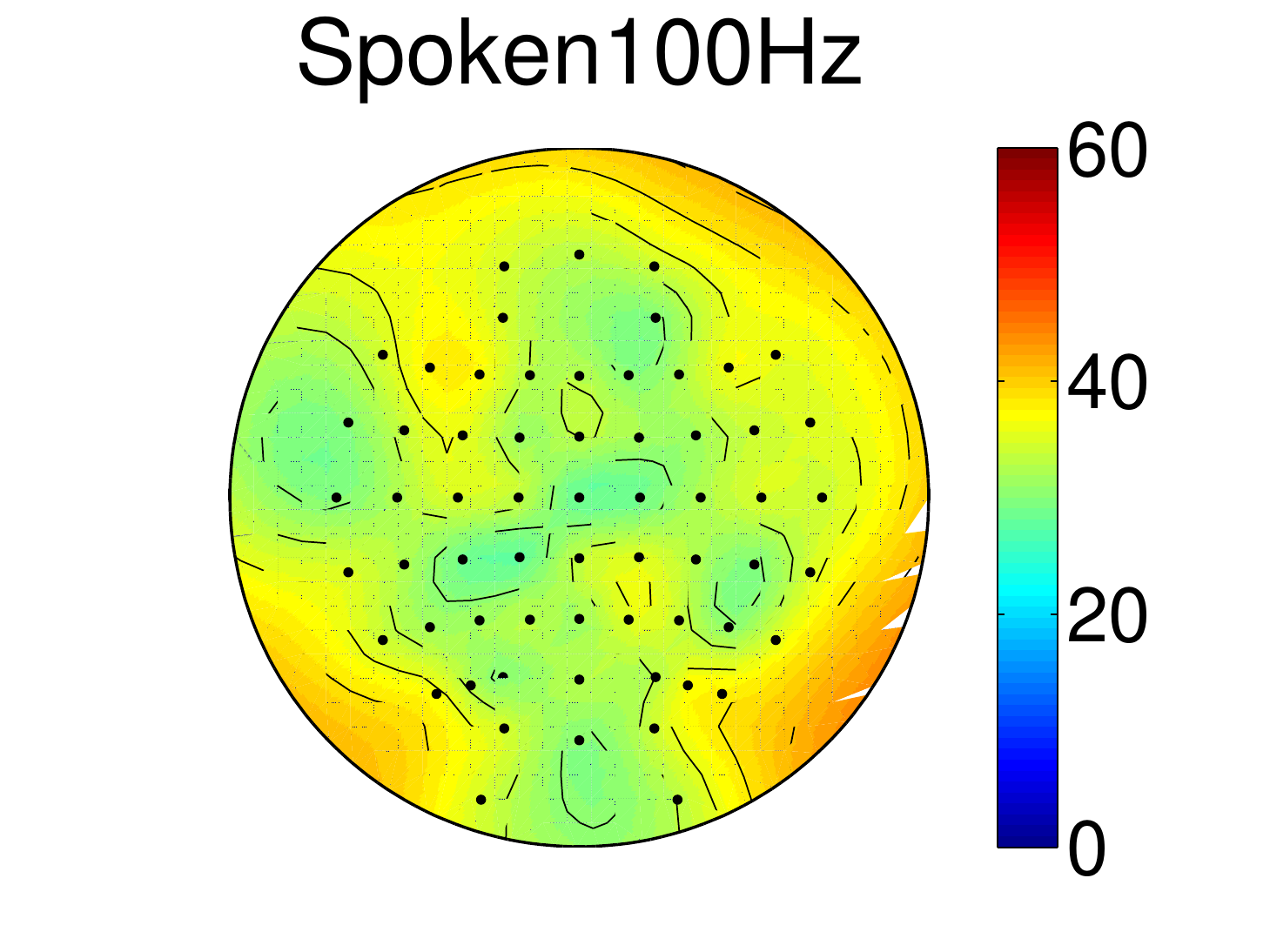}}
\hspace{-0.8em}
\subfigure{\includegraphics[scale=0.24]{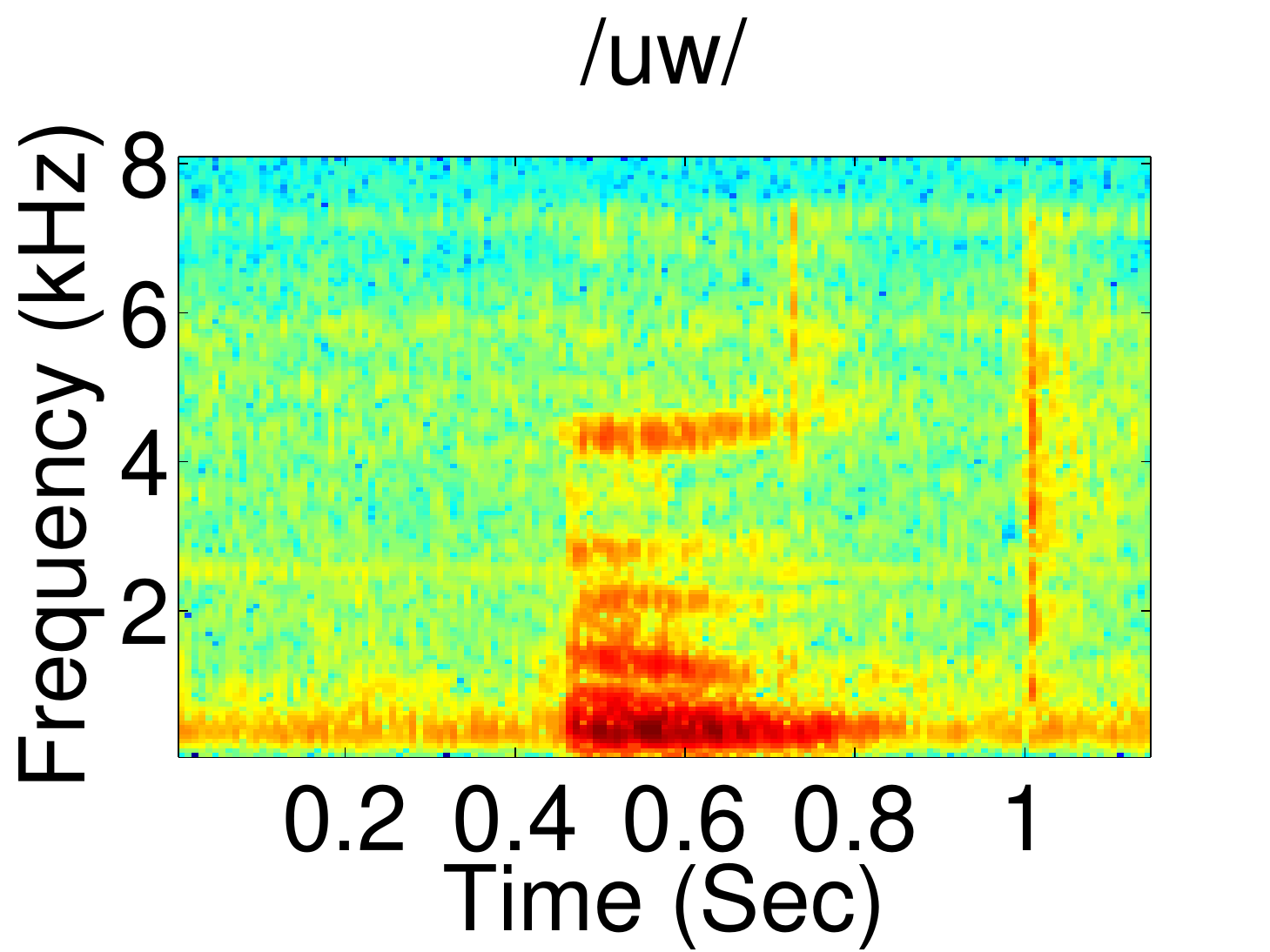}}
\hspace{-0.8em}
\subfigure{\includegraphics[scale=0.24]{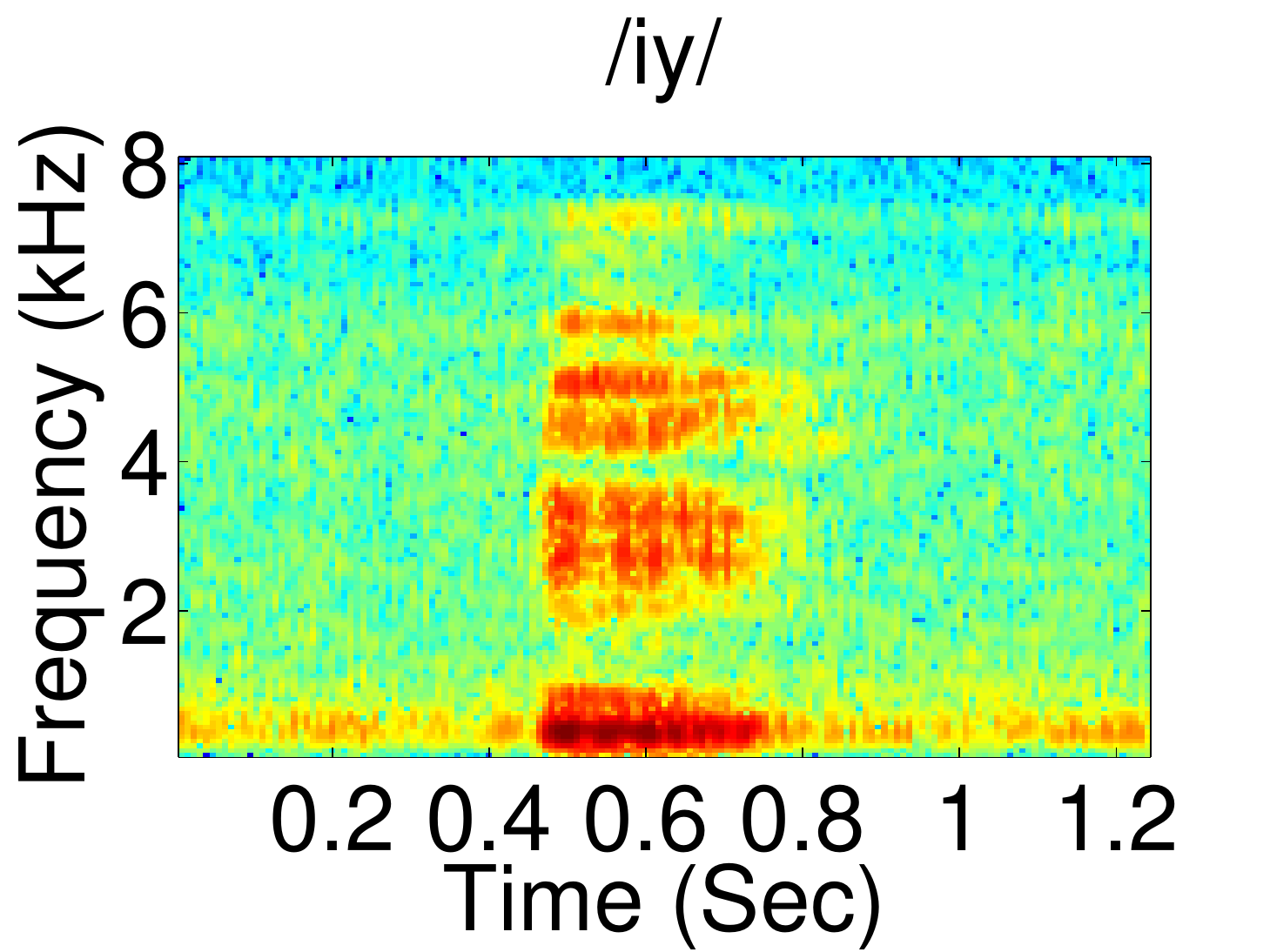}}
\hspace{-0.8em}
\subfigure{\includegraphics[scale=0.24]{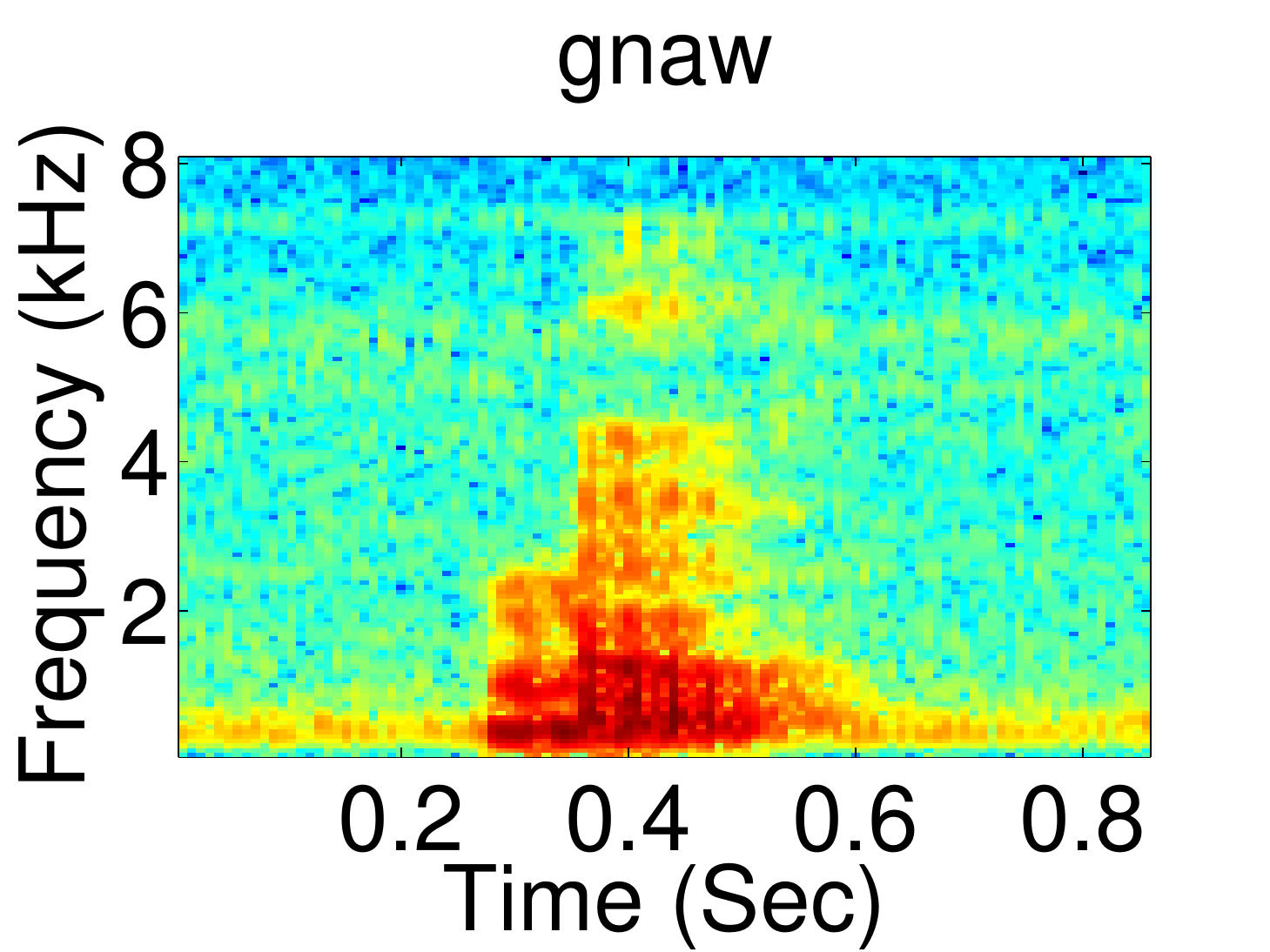}}
\hspace{-0.8em}
\subfigure{\includegraphics[scale=0.24]{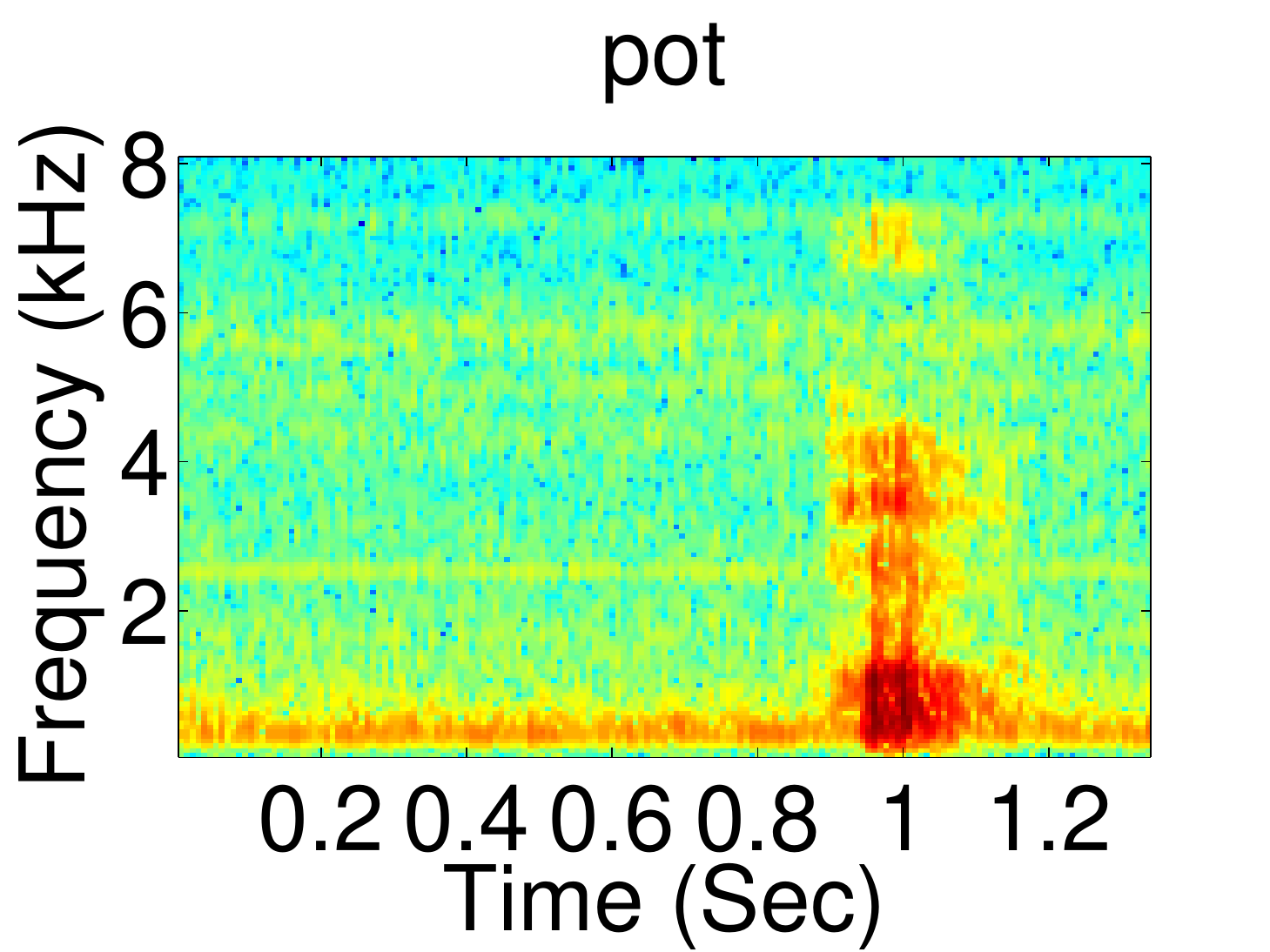}}
\hspace{-0.8em}
\subfigure{\includegraphics[scale=0.24]{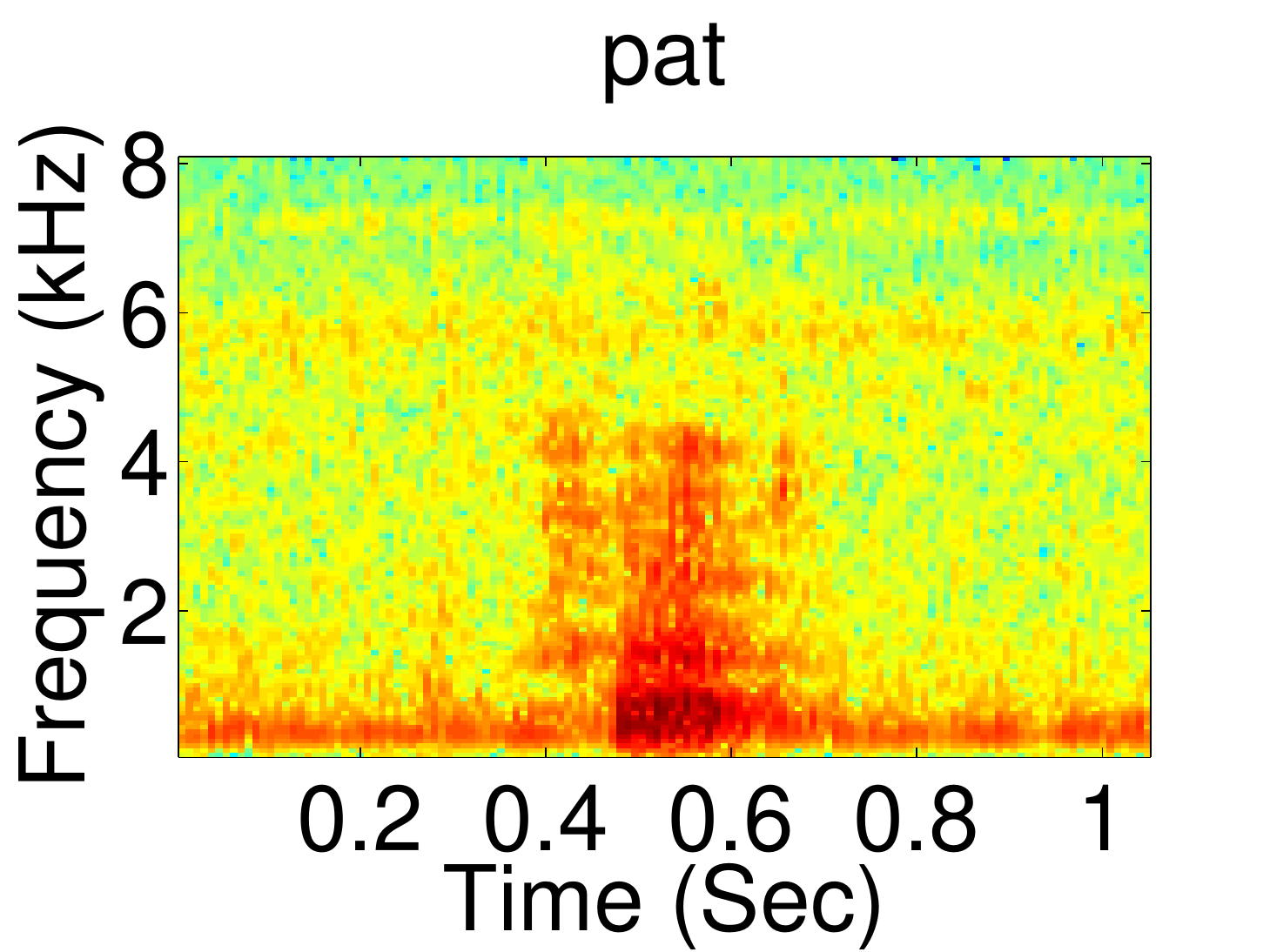}}
\vspace{-1mm}
\caption{ The 100Hz frequency components of imagined and spoken EEG scalp distribution (the scalp maps) and their corresponding speech spectrogram for ten phonemes, including $/n/$, $/tiy/$, $/piy/$, $/m/$, $/diy/$, $/uw/$, $/iy/$, $gnaw$, $pot$, and $pat$. The color bar for scalp distribution has been adjusted to the scope [0, 60]. All the phonemes and imagined-spoken EEG signals are randomly selected from participant MM05.  
}
\vspace{-3mm}
\label{scalpd}
\end{figure*} 

\subsection{Training process}
All proposed NES models are trained using following hyperparameters: all 'context' channel matrices use a weight decay of 1.0 $\times$ 10$^{-4}$ while EEG representations use a weight decay of 1.0 $\times$ 10$^{-5}$. All other weight matrices, including the Gaussian RBM filters, use a weight decay of 1.0 $\times$ 10$^{-4}$. In each EEG state, the batch size is set as 2000 and an initial learning rate is 0.1. Our NES-G model uses an initial learning rate of 0.02. Initial momentum is set as 0.5 and is increased linearly to 0.9 over 30 epochs. The EEG representation matrices are initialized to the 50 dimensional features. From each participant, we draw 100 trials to obtain the training sets (i.e., 1400 trials in total), and each phoneme is averagely collected.   

To generate phonemes, we have two sets of experiments: first, 62 channel EEG data are used as inputs to be co-context for each other; second, 10 selected channel data are used as the inputs. The proposed NES models are learned by a joint supervised and unsupervised training process, in which the middle layer (i.e., Gaussian RBM) is unsupervised initialized, and then the parameter sets are updated by supervised backward training. Specifically, the two procedures are implemented stepwise. The linear transformation $F$, $M$ and $J$ are all updated based on the NN parameter back propagation. 

For classification experiments, an extra softmax layer is added into three NES models as shown in Fig.~\ref{threemodels}(d). Both binary and multiple classification tasks are implemented. Ten channel inputs are used to train the proposed NES models. To avoid the overlap between training and evaluation sets, we randomly select 50 trials for each phoneme labels (i.e., 11 phonemes) from 14 participants as the evaluation set.
\begin{figure*}[htb]
\vspace{-4mm}
\centerline{\includegraphics[scale=0.5]{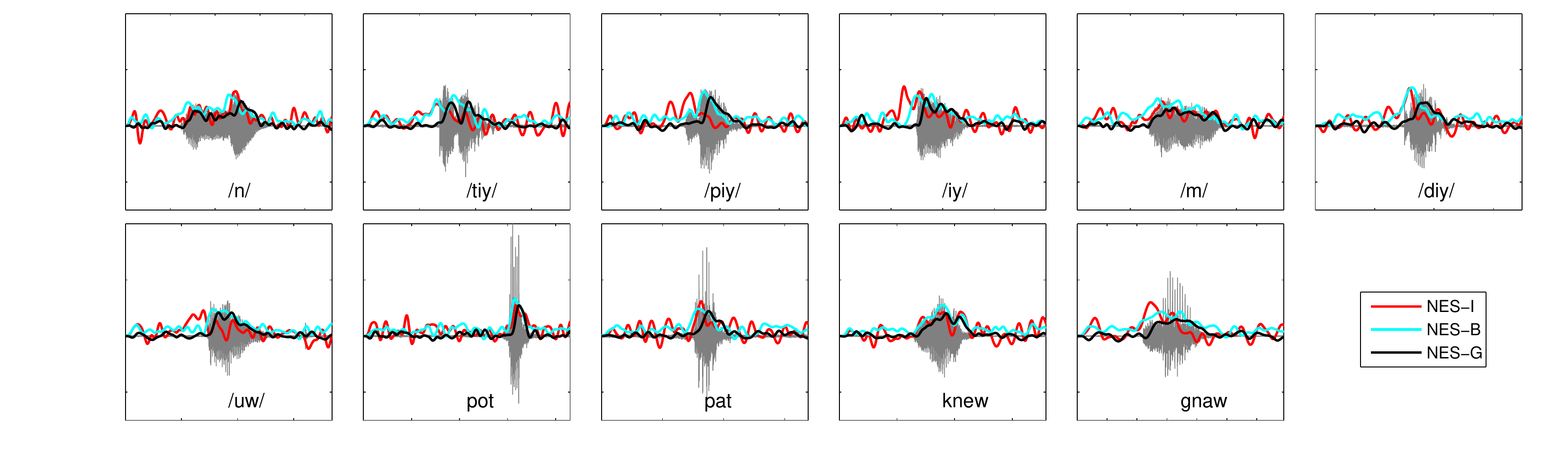}}
\vspace{-3mm}
\caption{The original speech components (gray) and the recovery speech envelops by three proposed NES models. Eleven different phonemes, including $/uw/$, $/iy/$, $pat$, and $knew$, are used, and all 62 channels EEG data recorded on a single participant (MM05) are used for training the NES models.}
\label{62channel}
\end{figure*}

\begin{figure*}[htb]
\vspace{-4mm}
\centerline{\includegraphics[scale=0.5]{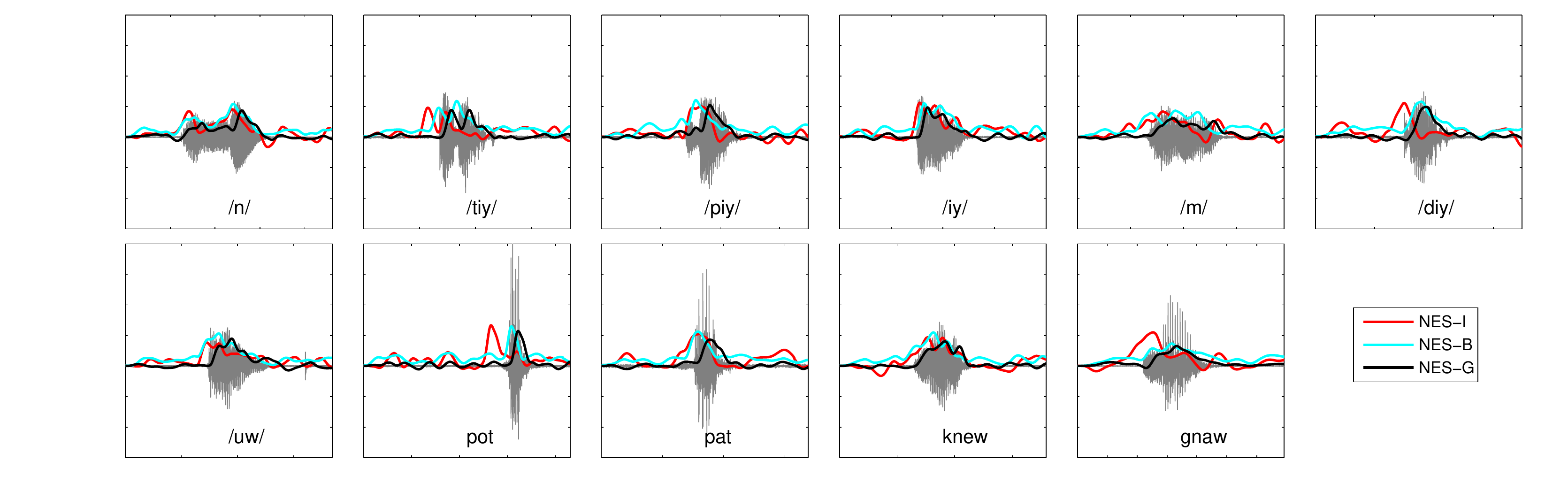}}
\vspace{-3mm}
\caption{The original speech components and the recovery speech envelops by three proposed NES models. Eleven different phonemes are used, and 10 selected EEG channels data recorded on a single participant (MM05) are used for training the NES models.}
\label{10channel}
\end{figure*}

\subsection{Imagined EEG and Spoken EEG}
In this study, we assume that there exists correlation between the imagined EEG and spoken EEG signals. To evident this proposition, we investigate the spectrum based scalp distribution of two types of EEG signals. 

As shown in Fig.~\ref{scalpd}, ten phonemes and their corresponding EEG signals are displayed in vertical array. The observations indicate that there exists unique pattern for each EEG-speech pair. For phoneme $/n/$, both imagined and spoken EEG signals show quite similar patterns in the central area of scalp, where is believed to be the speech related brain area. Other imagined-spoken EEG signals corresponding to phonemes, including $/tiy/$, $/piy/$, $/m/$, $/diy/$, $/iy/$, $gnaw$, and $pat$, also lead to the similar conclusion. In addition, the comparison between the phonemes with similar specgrogram distribution further shows that the corresponding imagined-spoken EEG signals are quite close to each other. For instance, the spectrograms of $/tiy/$ and $/piy/$ are similar, and their EEG patterns show comparable distributions. This phenomenon can also be found in the pair (i.e, $/m/$-$/n/$, and $pot$-$pat$).

All these intuitive comparisons demonstrate the effectiveness of introducing imagined-spoken EEG signals pair to strength EEG-speech recognition. Revealing the similarity of the two types of EEG signals benefits partial EEG signals based speech recognition and generation. Moreover, considering that the spoken EEG signals are directly correlated with the articulation procedure, the end-to-end NES system can also be applied to produce speech. 

\subsection{Recovery of Phoneme}
The trained end-to-end NES models can be used to generate speech signals based on the input EEG signals. Figure \ref{62channel} shows the recoveries of 11 different phonemes, including seven typical phonemic prompts and four phonetically-similar pairs, from 62 channels EEG signals recorded on a single participant (MM05). The output signal $\hat{\mathbf{h}}$ is regarded as the recovery of the speech envelope. Both speech signals and network output $\hat{\mathbf{h}}$ are normalized. 

As shown in Fig.~\ref{62channel}, all three proposed NES models can reflect the major fluctuations of the speech components. Specifically, NES-G model generally performs better than the other two models. The advantage of NES-G mainly includes it achieves smooth reconstruction of speech envelope. Especially, NES-G also demonstrates higher accuracy on picking out the major speech transitions compared with NES-I and NES-B models. However, all three models present some artifacts when recovering the envelopes. 

Figure \ref{10channel} shows the recoveries of all eleven phonemes obtained from 10 selected channels (i.e., FC6, FT8, C5, CP3, P3, T7, CP5, C3, CP1, and C4), which have higher correlations with speech activities \cite{zhao2015classifying}. These selected channels based experiment results confirm the involvement of selected regions during the planning of speech articulation \cite{zhao2015classifying}.

\begin{figure}[htb]
\vspace{-4mm}
\centerline{\includegraphics[scale=0.5]{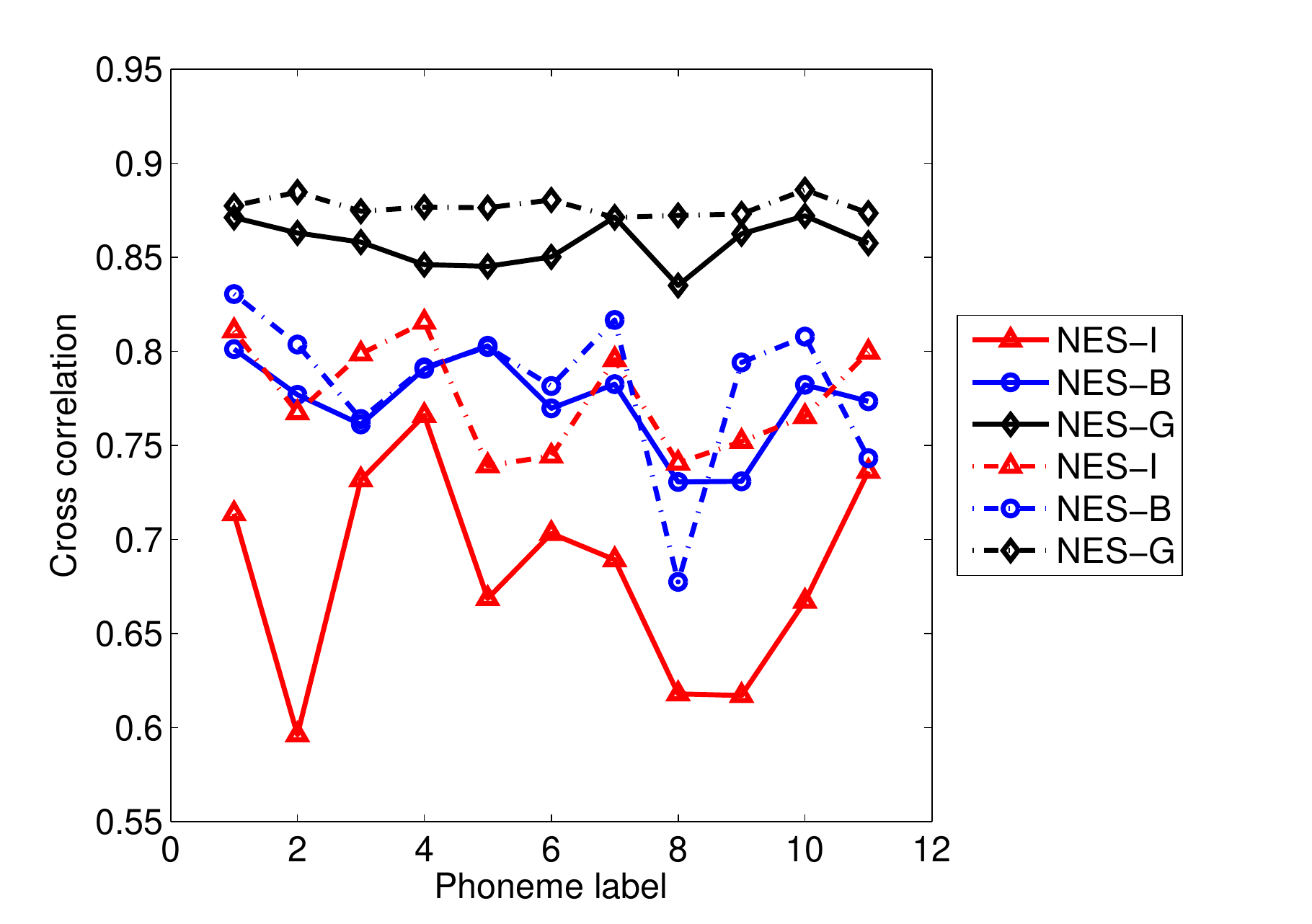}}
\vspace{-3mm}
\caption{Cross correlation coefficients of recovery phoneme-speech envelope pair for 11 phonemes. $-$ refers to 62 channel trained model, and $-.$ refers to 10 channel trained model.}
\label{Cross}
\end{figure}

In Fig.~\ref{10channel}, all three proposed NES models significantly improve the recoveries of the speech components compared with 62-channel based speech recovering. The intuitive envelope plotting shows that the 10-channel based speech envelopes are much smoother than 62-channel based speech envelopes. The potential explanation is that, the less-correlated EEG signals have been excluded from the computations, and accordingly the artifacts are reduced greatly. For $/iy/$, all three proposed NES models recover the speech component very well. By calculating the cross-correlation coefficients between the reconstructed phoneme and the speech envelope (as shown in Fig.\ref{Cross}), we can find that the NES-G model is slightly better than the NES-B model. Both NES-B and NES-G gain advantages against the NES-I model with respect to the envelope recovering and transitions catching. The results indicate that the gated network structure provides a stronger constraint to eliminate useless information.

\subsection{Classification Results}
\subsubsection{Binary Classification}
Due to highly mixed EEG signals, it is not easy to obtain an EEG-Phoneme pairwise  classification or mapping. Therefore, most previous studies are focused on binary classification \cite{zhao2015classifying}. As a comparison with the works on KARA ONE dataset, the proposed NES models are implemented to realize the binary phonological classification. Five binary classification tasks: vowels-only vs consonant ($\pm$ C/V), presence of nasal ($\pm$ Nasal), presence of bilabial ($\pm$ Bilab), presence of high-front vowel ($/iy/$), and presence of high-back vowel ($/uw/$), are conducted. The evaluation results are shown in Table.\ref{binaryc}. 
\begin{table}[ht]
\caption{ Accuracies of binary classification for five algorithms.} 
\centering 
\begin{threeparttable}
\scalebox{1.1}{
\begin{tabular*}{0.4\textwidth}{ c  c  c  c  c  c} 
\hline 
         & $\pm$ C/V  & $\pm$ Nasal & $\pm$ Bilab  & /iy / & /uw/   \\
\hline 
  INES        &0.25  &0.47  &0.53  &0.53  &0.74    \\
  IANES-B     &0.27  &0.59  &0.52  &0.62  &0.78     \\
  IANES-G     &0.41  &0.74  &0.71  &0.76  &0.87     \\
  SVM$^{*}$   &0.18  &0.64  &0.57  &0.59  &0.79       \\ 
  DBN$^{*}$   &0.87$^{\dagger}$  & $--$    &$--$    &$--$   &0.82$^{\dagger}$       \\ 
\hline 
\end{tabular*}
}
\end{threeparttable}
\vspace{-3mm}
\label{binaryc} 
\end{table}

In Table.\ref{binaryc}, $*$ indicates that the baseline results are collected from previous work \cite{zhao2015classifying}. In addition, the classification results of DBN on $\pm C/V$ and $\pm Nasal$ (labeled by $\dagger$) are obtained based on multi-modal features (i.e., facial, audio, and EEG features). The symbol $--$ indicates that the data are not available. 

Results show that the NES-G model demonstrates the best performance with respect to all five binary classification tasks, compared with nonlinear support vector machine (SVM) algorithm. Compared with DBN based algorithm, our proposed method achieves a slight improvement on $/uw/$ classification. Considering that the facial features can greatly help binary classification of $\pm C/V$ (vowel is with open vocal tract, while consonant is not), it is reasonable that DBN obtained a relative high result when compared with other EEG based classification approach.

\subsubsection{Multiple Classification}
We extend the EEG based speech classification into a multi-categories scenario. Compared with binary classification, this pairwise phonological mapping is more useful for accurate speech recognition and reconstruction, and is also more challenge.

\begin{figure}[!hbt]
\vspace{-4mm}
\centerline{\includegraphics[scale=0.7]{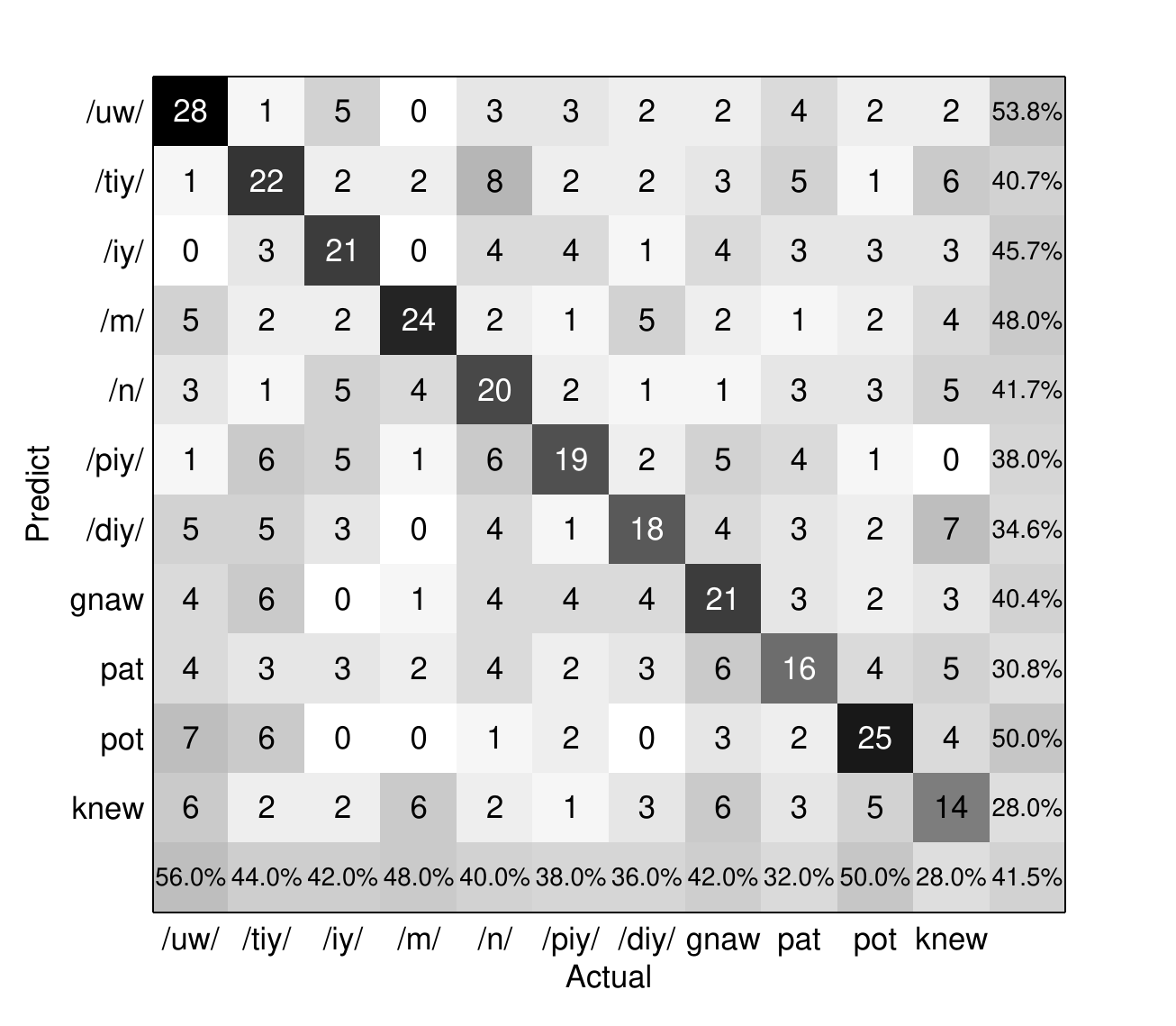}}
\vspace{-3mm}
\caption{The confusion matrix of eleven phonemes extracted from 14 participants in KARA ONE dataset. For each phoneme, 50 samples are randomly selected from the trials and then fed into the cross-trained NES-G model. The output is the predicted categories, when comparing with the actual categories.}
\label{confusionmatrix}
\end{figure}

Figure.~\ref{confusionmatrix} shows the confusion matrix of eleven phonemes. The diagonal cells show the number of correct classifications by the NES-G model. For example, 28 $/uw/$ trials are correctly classified as its original category. The horizontal axis represent the predicted classes. For instance, out of 52 $/uw/$ trial predictions, 53.8$\%$ are correct. Meanwhile, the actual classes are presented along the vertical axis. For example, out of 50 $/uw/$ cases, 56$\%$ are correctly predicted and 44$\%$ are predicted as the rest categories. In addition, this confusion matrix reveals that overall 41.5$\%$ of the predictions are correct by NES-G model. The classification results also show that the wrongly classified phonemes are very close to each other. The potential reason may lie that  the selected phonemes are quite similar to each other, for example, $/uw/$ and $/m/$ have close acoustic form. The proposed model may achieve better performance if it is trained by more diverse EEG-speech dataset.   
\begin{center}
\begin{table}[ht]
\vspace{-3mm}
\caption{The classification results for 11 phonemes by four approaches, including three proposed NES models and SVM.} 
\centering 
\begin{threeparttable}
\scalebox{1.2}{
\begin{tabular*}{0.35\textwidth}{ c  c  c  c  c} 
\hline 
  & NES-I  & NES-B & NES-G  & SVM$^{multi}$  \\
\hline 
  /uw/          &0.47  &0.53  &0.58  &0.24    \\
  /tiy/         &0.28  &0.37  &0.43  &0.21     \\
  /iy /          &0.34  &0.39  &0.41  &0.19     \\
  /m/           &0.35  &0.31  &0.51  &0.25    \\
  /n/           &0.27  &0.42  &0.39  &0.21     \\
  /piy/         &0.29  &0.31  &0.40  &0.18     \\
  /diy/         &0.39  &0.36  &0.39  &0.16      \\
  gnaw            &0.41  &0.46  &0.45  &0.23      \\
   pat            &0.32  &0.40  &0.41  &0.17      \\
   pot            &0.35  &0.39  &0.46  &0.27       \\
   knew           &0.31  &0.35  &0.33  &0.19    \\
\hline 
\end{tabular*}
}
\end{threeparttable}
\vspace{-3mm}
\label{classification} 
\end{table}
\end{center}

A comprehensive evaluation is also conducted with our proposed three NES models and SVM baseline in this study. A variant of SVM used for multi-class formulation is tested, given EEG feature vectors $x_{i}$ and $x_{j}$ with phoneme labels. This multi-class SVM supports linear kernels only. To enhance the performance of SVM, as presented in the work \cite{zhao2015classifying}, we compute various features over each window, such as the mean, median, standard deviation, variance, maximum, minimum, sum, spectral entropy, energy, skewness, and kurtosis. Moreover, the first and second derivatives are also calculated.  
 
The classification results have been summarized in Table.~\ref{classification}. The NES-G model achieves the best performance over the two other NES models. It may be explained as that, there exists strong correlation between the spoken-EEG signals and imagined EEG-signals. Therefore, incorporating the spoken-EEG as either bias or gate factor can both lead to an significant improvement on imagined-EEG feature extraction. This strong correlation may come from the similarity among the phonemes in KARA ONE dataset. The baseline SVM$^{multi}$ shows weak performance since these shallow features cannot reflect the correlations of EEG-Speech pairs. One can easily find that, even comparing with NES-I model, SVM$^{multi}$ still has a large gap on the classification results.

\section{Conclusion}
In this study, three end-to-end NES models are proposed. A multimodal fusion framework is developed aiming to translate imagined-EEG signals to the corresponding speech phonemes. The concept of joint-context tuple is introduced to abstract the EEG pattern as a probability distribution. Unlike conventional feature extraction approaches, the proposed NES models incorporate the representation of EEG signals into the NN structure, in which EEG signals are projected into a feature space that is learned by the NN through both supervised and unsupervised training. In the two augmented NES models, referring as NES-B and NES-G, the spoken EEG signals are included to condition imagined EEG as a bias channel or gate channel. In addition, the speech phonemes are also projected into the same feature space to realize the modal fusion. The experimental results show that the proposed NES models can recover speech envelopes well based on the input EEG signals. The EEG based speech classification results indicate that the proposed NES models outperform a baseline method SVM$^{multi}$ for 11 phonological categories. In the binary classification task, the proposed NES-G model overall achieves the best performance than nonlinear SVM classifier. Specifically, NES-G model averagely obtains the best classification results compared with NES-I and NES-B models.   

\ifCLASSOPTIONcaptionsoff
  \newpage
\fi

\bibliographystyle{IEEEtran}
\bibliography{references}

\begin{thebibliography}{10}
\providecommand{\url}[1]{#1}
\csname url@samestyle\endcsname
\providecommand{\newblock}{\relax}
\providecommand{\bibinfo}[2]{#2}
\providecommand{\BIBentrySTDinterwordspacing}{\spaceskip=0pt\relax}
\providecommand{\BIBentryALTinterwordstretchfactor}{4}
\providecommand{\BIBentryALTinterwordspacing}{\spaceskip=\fontdimen2\font plus
\BIBentryALTinterwordstretchfactor\fontdimen3\font minus
  \fontdimen4\font\relax}
\providecommand{\BIBforeignlanguage}[2]{{%
\expandafter\ifx\csname l@#1\endcsname\relax
\typeout{** WARNING: IEEEtran.bst: No hyphenation pattern has been}%
\typeout{** loaded for the language `#1'. Using the pattern for}%
\typeout{** the default language instead.}%
\else
\language=\csname l@#1\endcsname
\fi
#2}}
\providecommand{\BIBdecl}{\relax}
\BIBdecl

\bibitem{toda2008statistical}
T.~Toda, A.~W. Black, and K.~Tokuda, ``Statistical mapping between articulatory
  movements and acoustic spectrum using a gaussian mixture model,''
  \emph{Speech Communication}, vol.~50, no.~3, pp. 215--227, 2008.

\bibitem{di2015low}
G.~M. Di~Liberto, J.~A. O’Sullivan, and E.~C. Lalor, ``Low-frequency cortical
  entrainment to speech reflects phoneme-level processing,'' \emph{Current
  Biology}, vol.~25, no.~19, pp. 2457--2465, 2015.

\bibitem{lotte2007review}
F.~Lotte, M.~Congedo, A.~L{\'e}cuyer, F.~Lamarche, and B.~Arnaldi, ``A review
  of classification algorithms for eeg-based brain--computer interfaces,''
  \emph{Journal of neural engineering}, vol.~4, no.~2, p.~R1, 2007.

\bibitem{gao2014visual}
S.~Gao, Y.~Wang, X.~Gao, and B.~Hong, ``Visual and auditory brain--computer
  interfaces,'' \emph{IEEE Transactions on Biomedical Engineering}, vol.~61,
  no.~5, pp. 1436--1447, 2014.

\bibitem{herff2016automatic}
C.~Herff and T.~Schultz, ``Automatic speech recognition from neural signals: A
  focused review,'' \emph{Frontiers in Neuroscience}, vol.~10, 2016.

\bibitem{matsumoto2014classification}
M.~Matsumoto and J.~Hori, ``Classification of silent speech using support
  vector machine and relevance vector machine,'' \emph{Applied Soft Computing},
  vol.~20, pp. 95--102, 2014.

\bibitem{sun2016enhanced}
P.~Sun and J.~Qin, ``Enhanced factored three-way restricted boltzmann machines
  for speech detection,'' \emph{arXiv preprint arXiv:1611.00326}, 2016.

\bibitem{suppes1997brain}
P.~Suppes, Z.-L. Lu, and B.~Han, ``Brain wave recognition of words,''
  \emph{Proceedings of the National Academy of Sciences}, vol.~94, no.~26, pp.
  14\,965--14\,969, 1997.

\bibitem{dasalla2009single}
C.~S. DaSalla, H.~Kambara, M.~Sato, and Y.~Koike, ``Single-trial classification
  of vowel speech imagery using common spatial patterns,'' \emph{Neural
  Networks}, vol.~22, no.~9, pp. 1334--1339, 2009.

\bibitem{d2009toward}
M.~D’Zmura, S.~Deng, T.~Lappas, S.~Thorpe, and R.~Srinivasan, ``Toward eeg
  sensing of imagined speech,'' in \emph{International Conference on
  Human-Computer Interaction}.\hskip 1em plus 0.5em minus 0.4em\relax Springer,
  2009, pp. 40--48.

\bibitem{lopez2012auditory}
M.~Lopez-Gordo, E.~Fernandez, S.~Romero, F.~Pelayo, and A.~Prieto, ``An
  auditory brain--computer interface evoked by natural speech,'' \emph{Journal
  of neural engineering}, vol.~9, no.~3, p. 036013, 2012.

\bibitem{porbadnigk2009eeg}
A.~Porbadnigk, M.~Wester, and T.~S. Jan-p Calliess, ``Eeg-based speech
  recognition impact of temporal effects,'' 2009.

\bibitem{hausfeld2012pattern}
L.~Hausfeld, F.~De~Martino, M.~Bonte, and E.~Formisano, ``Pattern analysis of
  eeg responses to speech and voice: Influence of feature grouping,''
  \emph{Neuroimage}, vol.~59, no.~4, pp. 3641--3651, 2012.

\bibitem{wang2013analysis}
L.~Wang, X.~Zhang, X.~Zhong, and Y.~Zhang, ``Analysis and classification of
  speech imagery eeg for bci,'' \emph{Biomedical Signal Processing and
  Control}, vol.~8, no.~6, pp. 901--908, 2013.

\bibitem{zhao2015classifying}
S.~Zhao and F.~Rudzicz, ``Classifying phonological categories in imagined and
  articulated speech,'' in \emph{2015 IEEE International Conference on
  Acoustics, Speech and Signal Processing (ICASSP)}.\hskip 1em plus 0.5em minus
  0.4em\relax IEEE, 2015, pp. 992--996.

\bibitem{o2015attentional}
J.~A. O'Sullivan, A.~J. Power, N.~Mesgarani, S.~Rajaram, J.~J. Foxe, B.~G.
  Shinn-Cunningham, M.~Slaney, S.~A. Shamma, and E.~C. Lalor, ``Attentional
  selection in a cocktail party environment can be decoded from single-trial
  eeg,'' \emph{Cerebral Cortex}, vol.~25, no.~7, pp. 1697--1706, 2015.

\bibitem{yoshimura2016decoding}
N.~Yoshimura, A.~Nishimoto, A.~N. Belkacem, D.~Shin, H.~Kambara, T.~Hanakawa,
  and Y.~Koike, ``Decoding of covert vowel articulation using
  electroencephalography cortical currents,'' \emph{Frontiers in neuroscience},
  vol.~10, 2016.

\bibitem{ngiam2011multimodal}
J.~Ngiam, A.~Khosla, M.~Kim, J.~Nam, H.~Lee, and A.~Y. Ng, ``Multimodal deep
  learning,'' in \emph{Proceedings of the 28th international conference on
  machine learning (ICML-11)}, 2011, pp. 689--696.

\bibitem{srivastava2012multimodal}
N.~Srivastava and R.~R. Salakhutdinov, ``Multimodal learning with deep
  boltzmann machines,'' in \emph{Advances in neural information processing
  systems}, 2012, pp. 2222--2230.

\bibitem{socher2014grounded}
R.~Socher, A.~Karpathy, Q.~V. Le, C.~D. Manning, and A.~Y. Ng, ``Grounded
  compositional semantics for finding and describing images with sentences,''
  \emph{Transactions of the Association for Computational Linguistics}, vol.~2,
  pp. 207--218, 2014.

\bibitem{frome2013devise}
A.~Frome, G.~S. Corrado, J.~Shlens, S.~Bengio, J.~Dean, T.~Mikolov
  \emph{et~al.}, ``Devise: A deep visual-semantic embedding model,'' in
  \emph{Advances in neural information processing systems}, 2013, pp.
  2121--2129.

\bibitem{mnih2007three}
A.~Mnih and G.~Hinton, ``Three new graphical models for statistical language
  modelling,'' in \emph{Proceedings of the 24th international conference on
  Machine learning}.\hskip 1em plus 0.5em minus 0.4em\relax ACM, 2007, pp.
  641--648.

\bibitem{kiros2014multimodal}
R.~Kiros, R.~Salakhutdinov, and R.~S. Zemel, ``Multimodal neural language
  models.'' in \emph{ICML}, vol.~14, 2014, pp. 595--603.

\bibitem{yamashita2014bernoulli}
T.~Yamashita, M.~Tanaka, E.~Yoshida, Y.~Yamauchi, and H.~Fujiyoshi, ``To be
  bernoulli or to be gaussian, for a restricted boltzmann machine.'' in
  \emph{ICPR}, 2014, pp. 1520--1525.

\bibitem{memisevic2010learning}
R.~Memisevic and G.~E. Hinton, ``Learning to represent spatial transformations
  with factored higher-order boltzmann machines,'' \emph{Neural Computation},
  vol.~22, no.~6, pp. 1473--1492, 2010.

\bibitem{nguyen2013learning}
T.~D. Nguyen, T.~Tran, D.~Q. Phung, and S.~Venkatesh, ``Learning parts-based
  representations with nonnegative restricted boltzmann machine.'' in
  \emph{ACML}, 2013, pp. 133--148.

\end{thebibliography}

\end{document}